\documentclass[graybox]{svmult}

\usepackage{mathptmx}       
\usepackage{helvet}         
\usepackage{courier}        
\usepackage{type1cm}        
\usepackage{makeidx}         
\usepackage{graphicx}        
\usepackage{multicol}        
\usepackage[bottom]{footmisc}

\usepackage{etex, xy}
\xyoption{all}

\usepackage{amsmath,amsbsy,amssymb}

\usepackage{color}        

\newcommand{\pisiSE}{$\Pi\Sigma^*$}

\newcommand{\KK}{\mathbb{K}}
\newcommand{\NN}{\mathbb{N}}
\newcommand{\ZZ}{\mathbb{Z}}
\newcommand{\FF}{\mathbb{F}}
\newcommand{\HF}{\mathbb{H}}
\newcommand{\QQ}{\mathbb{Q}}

\newcommand{\dfield}[2]{({#1},{#2})}
\newcommand{\SigmaP}{\texttt{Sigma}}
\let\set\mathbb
\newcommand{\Dconst}[2]{{\rm const}_{#2}{#1}}
\newcommand{\SH}{{\mathcal S}}
\newcommand{\fct}[3]{{#1:#2 \to #3}}
\newcommand{\ev}{\mathrm{ev}}
\newcommand{\vect}[1]{\vec{#1}}

\newcommand{\seqR}{{S}(\KK)}
\newcommand{\Shift}{{\mathcal S}}

\makeindex             

\let\set\mathbb

\newcommand{\ep}{\varepsilon}

\newcounter{Cmmacnt}
\def\restartmma{\setcounter{Cmmacnt}{0}}
\restartmma \catcode`|=\active
\def|#1|{\mathrm{#1}}
\catcode`|=12
\newenvironment{Cmma}{
 \par\smallskip
 \catcode`|=\active
 \parskip=0pt\parindent=0pt 
 \footnotesize
 \def\CIn##1\\{%
   \def\linebreak{\hfill\break\null\qquad}%
   \refstepcounter{Cmmacnt}
   \hangindent=2.5em\hangafter=0
   \leavevmode
   \llap{\tiny\sffamily In[\arabic{Cmmacnt}]:=\kern.5em}%
   \mathversion{bold}\scriptsize$\tt\bf\displaystyle##1$\normalsize
   \mathversion{normal}
 }%
 \def\CPrint##1\\{%
   \def\linebreak{$\hfill\break\null\hfill$}%
   \kern\abovedisplayskip\par
   \hangindent=2.5em\hangafter=0
   \leavevmode
   \llap{}
   \scriptsize$\displaystyle\tt##1$\normalsize\hfill\null\par
   \kern\belowdisplayskip
 }%
 \def\COut##1\\{%
   \def\linebreak{$\hfill\break\null\hfill$}%
   \kern\abovedisplayskip\par
   \hangindent=2.5em\hangafter=0
   \leavevmode
   \llap{\tiny\sffamily Out[\arabic{Cmmacnt}]=\kern.5em}
   \scriptsize$\displaystyle\tt##1$\normalsize\hfill\null\par
   \kern\belowdisplayskip\vspace*{-0.3cm}
 }%
 \def\CWarning##1##2\\{%
   \def\linebreak{\hfill\break}%
   \hangindent=2.5em\hangafter=0
   \leavevmode
   {\scriptsize##1 : ##2}\par}%
}{%
 \par\smallskip
}

\newcommand{\CmyOut}[1]{{\small\sffamily Out[#1]}}

\def\CMLabel#1{{\refstepcounter{Cmmacnt}\label{#1}}\addtocounter{Cmmacnt}{-1}}

\newcommand{\MyFrame}[1]{
\hspace*{-0.08cm}\fcolorbox[rgb]{0,0,0}{1,1,1}{
\begin{minipage}{11.31cm}#1\end{minipage}}
}

\begin{document}

\title*{Simplifying Multiple Sums in Difference Fields}
\author{Carsten Schneider}
\institute{Carsten Schneider \at Research Institute for Symbolic Computation (RISC), Johannes Kepler University Linz, \email{Carsten.Schneider@risc.jku.at}}
%
%
\maketitle

\vspace*{-0.4cm}

\abstract{
In this survey article we present difference field algorithms for symbolic summation. Special emphasize is put on new aspects in how the summation problems are rephrased in terms of difference fields, how the problems are solved there, and how the derived results in the given difference field can be reinterpreted as solutions of the input problem. 
The algorithms are illustrated with the Mathematica package \SigmaP\ by discovering and proving new harmonic number identities extending those from (Paule and Schneider, 2003). In addition, the newly developed package \texttt{EvaluateMultiSums} is introduced that combines the presented tools. In this way, large scale summation problems for the evaluation of Feynman diagrams in QCD (Quantum ChromoDynamics) can be solved completely automatically.}

\section{Introduction}

We will elaborate a
symbolic summation toolbox based on up--to-date algorithms in the setting of difference fields. It contains hypergeometric and $q$--hypergeometric summation, see, e.g., ~\cite{Gosper:78,Zeilberger:91,Petkov:92,AequalB,Paule:95,PauleSchorn:95,vanHoeij:99,Koepf:12,CK:12} and~\cite{Koornwinder:93,PauleRiese:97,Bauer:99} respectively\index{summation!hypergeometric}\index{summation!$q$--hypergeometric}, and it can deal with multiple sums covering big parts of ($q$--)hypergeometric multi-summation\index{summation!multiple sum}~\cite{Wilf:92,Wegschaider,Riese:03,AZ:06} and ($q$--)holonomic sequences\index{summation!holonomic sequence}~\cite{Zeilberger:90a,Chyzak:00,Schneider:05d,Koutschan:2010}. 

This difference field approach started with Karr's theory of \pisiSE-fields and his indefinite summation algorithm~\cite{Karr:81,Karr:85}; for the continuous analogue of indefinite integration see~\cite{Risch:69}. In principle, the algorithm solves the telescoping problem in a given field generated by indefinite nested sums and products. In this article we restrict this input class to \textit{nested hypergeometric sum expressions}\index{summation!nested hypergeometric sum} (see Definition~\ref{Def:NestedHgSums}), i.e., expressions where the arising products represent hypergeometric sequences, and the sums and products occur only as polynomial expressions in the numerators; evaluating such expressions produces d'Alembertian sequences~\cite{Abramov:94},\cite{Petkov:2013}, a subclass of Liouvillian sequences~\cite{Singer:99}. We point out that exactly this restriction covers all the summation problems that have been relevant in practical problem solving so far. There we solve the following fundamental problem: given a nested hyperge\-ometric sum expression, calculate an alternative expression such that the occurring sums are algebraically independent~\cite{Schneider:10c}; for related work see~\cite{Bluemlein:04,Schneider:05c,Kauers:08,Singer:08,ABS:13}. In addition, the found representation should be given in terms of sums and products that are as simple as possible; for a general framework we refer to~\cite{PauleNemes:97}. This is possible by representing the sums and products in \pisiSE-fields reflecting certain optimality properties:  
We will exploit simplifications taking into account, e.g., the minimal nesting depth~\cite{Schneider:04a,Schneider:05f,Schneider:08c,Schneider:10b,Schneider:10a} or minimal degrees~\cite{Schneider:05c,Schneider:07d,Petkov:10}.

Besides indefinite summation, we aim at the transformation of a definite multiple sum to nested hypergeometric sums. As for the special case of hypergeometric summation~\cite{AequalB,Zeilberger:91} one looks for a recurrence of such a sum~\cite{Schneider:01}. If one succeeds, one computes all solutions of the found recurrence that are expressible in terms of nested hypergeometric sum expressions; for solvers of recurrences in terms of polynomials and \pisiSE-fields see~\cite{Petkov:92,Abramov:94,Singer:99} and~\cite{Bron:00,Schneider:01,Schneider:05a,ABPS:13}, respectively. Finally, one tries to combine the solutions to an expression that equals the input sum.

All these algorithms (also for the $q$-hypergeometric and mixed case) are available in the summation package Sigma~\cite{Schneider:07a}\index{Sigma, summation package} and have been used to discover and prove demanding identities from combinatorics or related fields, like, e.g., in~\cite{Schneider:06c,Schneider:07b,Schneider:07c,Schneider:09a}. 
A typical example is the sum
$$A_{\alpha}(a)=\sum_{k=0}^{a}\big(1+\alpha(n-2k)S_1(k)\big)\binom{n}{k}^{\alpha}\quad\text{with }S_1(k)=\sum_{i=1}^k\frac{1}{i}$$
\noindent which is connected to supercongruences of the Ap{\'e}ry numbers.
For the treatment of the cases $\alpha\in\{1,2,\dots,5\}$ and $\alpha>5$ we refer to~\cite{Schneider:03} and~\cite{Krattenthaler:04}, respectively. As running example we will discover and prove the following identities\footnote{For identities~\eqref{Id:AM1}, \eqref{Id:AM2} we point also to~\cite{Chu2005}; for their indefinite versions see~\eqref{Equ:A1Indef},\eqref{Equ:A2Indef} below.}
{\allowdisplaybreaks[3]
\begin{align}
A_{-1}(n)=&(n+1)S_1(n)+1,\label{Id:AM1}\\
A_{-2}(n)=&\frac{(n+1)^2}{(n+2)^2}+\frac{\left(n+2
   \left(n^2+3 n+2\right) S_1(n)+3\right)
   (n+1)}{(n+2)^2},\label{Id:AM2}\\
A_{-3}(n)=&(-1)^n(5S_{-3}(n)(n+1)^3-6S_{-2,1}(n) (n+1)^3)+6 S_1(n)(n+1)+1,\label{Id:AM3}\\
A_{-4}(n)=&\frac{(-1)^n\binom{2n}{n}^{-1}(n+1)^5}{(4 n
   (n+2)+3)}\left(\frac{7}{2} \sum
   _{i=1}^n \frac{(-1)^i \binom{2i}{i}}{i^3}-5 \sum _{i=1}^n
   \frac{(-1)^i \binom{2i}{i} S_1(i)}{i^2}\right)+\nonumber\\[-0.1cm]
&+\frac{(10 (n+1) S_1(n)+3)
   (n+1)}{2 n+3}\label{Id:AM4}
\end{align}}
\noindent 
where the harmonic sums\index{harmonic sum}~\cite{Bluemlein:99,Vermaseren:99} are defined by

\vspace*{-0.3cm}

\begin{equation}\label{Equ:HarmonicSums}
S_{m_1,\dots,m_k}(n)=
\sum_{i_1=1}^n\frac{\text{\small$\text{sign}(m_1)^{i_1}$}}{i_1^{|m_1|}}\dots
\sum_{i_k=1}^{i_{k-1}}\frac{\text{\small$\text{sign}(m_k)^{i_k}$}}{i_k^{|m_k|}},\quad\quad m_i\in\ZZ\setminus\{0\}.
\end{equation}
\noindent We emphasize that exactly this type of nested hypergeometric sums is related to summation problems coming from QCD like., e.g., in~\cite{Moch:04,BKKS:09} or in~\cite{Schneider:08e,HYP2,ABHKSW:12,BHKS:13}. More precisely, 2-- and 3--loop Feynman integrals\index{Feynman integral} with at most one mass and with operator insertion can be transformed to multiple sums\index{summation!multiple sum}~\cite{BKSF:12} depending on a discrete Mellin parameter $n$. Then these sums must be simplified in terms of special functions~\cite{AB:2013}, such as harmonic sums~\eqref{Equ:HarmonicSums}, their infinite versions of multiple zeta values\index{multiple zeta value}~\cite{MZV} and generalizations like $S$-sums\index{generalized harmonic sum}~\cite{Moch:02,ABS:13} and cyclotomic harmonic sums\index{cyclotomic harmonic sum}~\cite{ABS:11}. In recent calculations~\cite{ABFHKRRSW:12} also binomial sums as in~\eqref{Id:AM4} arose.
For certain sum classes we point to efficient tools like~\cite{Vermaseren:99,Moch:02}. For harder sums such as~\cite{ABHKSW:12}
\small
\begin{align}\nonumber
\text{\scriptsize$\displaystyle\sum_{j=0}^{n-3}$}&\text{\scriptsize$\displaystyle\sum_{k=0}^j \sum_{l=0}^k \sum_{q=0}^{-j+n-3} \sum_{s=1}^{-l+n-q-3} \sum_{r=0}^{-l+n-q-s-3}$}\tfrac{\binom{j+1}{k+1}
\binom{k}{l} \binom{n-1}{j+2} \binom{-j+n-3}{q} \binom{-l+n-q-3}{s}
\binom{-l+n-q-s-3}{r} r! (-l+n-q-r-s-3)! (s-1)!}{(-l+n-q-2)!(-j+n-1)(n-q-r-s-2) (q+s+1)}\label{Equ:QCDSum}\\[-0.1cm]
&(-1)^{-j+k-l+n-q-3}\Big[4 S_1(-j+n-1)-4S_1(-j+n-2)-2S_1(k)-(S_1(-l+n-q-2)\\[-0.1cm]
&\hspace*{2cm}+S_1(-l+n-q-r-s-3)-2 S_1(r+s))+2 S_1(s-1)-2S_1(r+s)\Big]\nonumber
\end{align}
\normalsize
\noindent the summation techniques under consideration work successfully and are applied automatically in the newly developed package \texttt{EvaluateMultiSums}~\cite{Schneider:12a}. In this way, millions of multiple sums~\cite{HYP2,BHKS:13} could be treated. In addition, the presented packages and algorithms are used by an enhanced version~\cite{ABHKSW:12,ABFHKRRSW:12} of the method of hyperlogarithms~\cite{Brown:09} and by new algorithms for the calculation of $\ep$-expansions~\cite{BKSF:12,ABRS:12} utilizing multi-summation and integration methods~\cite{Wegschaider,AZ:06,Schneider:05d}. 

The outline of this article is as follows. In Section~\ref{Sec:BasicMechanism} we present the basic mechanism how expressions in terms of indefinite nested sums and products can be rephrased in a difference field. For interested readers details are given in Section~\ref{Sec:DFDetails}. Readers, that are primarily interested in the summation tools and how they can be applied with the summation package \SigmaP, can jump directly to Section~\ref{Sec:SumTechniques}. Finally, in Section~\ref{Sec:EMS} the new package \texttt{EvaluateMuliSums} is introduced that combines all the presented summation methods. It enables one to simplify definite nested sums to indefinite nested hypergeometric sums completely automatically.

\vspace*{-0.3cm}

\section{Indefinite summation: the basic mechanism}\label{Sec:BasicMechanism}

We will work out the basic principles how indefinite summation can be carried out in the setting of difference fields.
This will be illustrated by the task to simplify

\vspace*{-0.2cm}

\begin{equation}\label{Equ:A1}
A_{-1}(a)=\sum_{k=0}^aF(k)=\sum_{k=0}^{a}\big(1-(n-2k)S_1(k)\big)\binom{n}{k}^{-1},
\end{equation}

\vspace*{-0.1cm}

\noindent to be more precise, by the task to solve the following problem.

\smallskip

\noindent\MyFrame{\noindent \textbf{Problem T: Telescoping.}\label{Problem:T} \textit{Given} a summand $F(k)$.
\textit{Find} an expression $G(k)$ such that

\vspace*{-0.5cm}

\begin{equation}\label{Equ:TeleSum}
G(k+1)=G(k)+F(k+1)
\end{equation}
\noindent and such that $G(k)$ is not ``more complicated'' than $F(k)$.}

\smallskip

\noindent For our given $F(k)$ in~\eqref{Equ:A1} we will compute for $k\geq0$ the solution
\begin{equation}\label{Equ:TeleSumSol}
G(k)=((k+1) S_{1}(k)+1)\tbinom{n}{k}^{-1}+c,\quad\quad c\in\QQ(n).
\end{equation}
Since $A_{-1}(a)$ and $G(a)$ satisfy both the recurrence
$A(a+1)=A(a)+F(a+1)$, they are equal for $a\geq0$ if they agree at $a=0$; this is the case with $c=0$. Hence we get for $A_{-1}(a)$, and with the same technique for $A_{-2}(a)$, the simplifications

\vspace*{-0.5cm}

\begin{align}\label{Equ:A1Indef}
A_{-1}(a)&=((a+1) S_{1}(a)+1)\text{\small$\binom{n}{a}^{-1}$},\\[-0.1cm]
A_{-2}(a)&=\frac{(n+1)^2}{(n+2)^2}+\frac{(a+1) (-a+2
   n+2 (a+1) (n+2) S_1(a)+3)}{(n+2)^2}\text{\small$\binom{n}{a}^{-2}$}.\label{Equ:A2Indef}
\end{align}

\vspace*{-0.15cm}

\noindent Note that for the special case $a=n$ this simplifies to~\eqref{Id:AM1} and~\eqref{Id:AM2}, respectively.

Subsequently, we give more details how this solution $G(k)$ for~\eqref{Equ:TeleSum} can be derived automatically. First observe that the occurring sums can be written in terms of indefinite sums and products: for all $k\in\NN$,

\vspace*{-0.3cm}

$$S_1(k)=\text{\small$\sum_{i=1}^k$}\tfrac{1}{i}\quad\text{ and }\quad\tbinom{n}{k}=\text{\small$\prod_{i=1}^k$}\tfrac{n-i+1}{k};$$

\vspace*{-0.1cm}

\noindent here $n$ is considered as a variable.
Now let $\SH_k$ be the shift operator w.r.t.\ $k$.
Then using the shift behavior of the summand objects, namely

\vspace*{-0.4cm}

\begin{equation}\label{Equ:SummandShiftBehavior}
\SH_k n=n,\quad
\SH_k k=k+1,\quad \SH_k\tbinom{n}{k}=\tfrac{n-k}{k+1}\tbinom{n}{k},\quad
\SH_k S_1(k)=S_1(k)+\tfrac{1}{k+1},\\
\end{equation}

\vspace*{-0.1cm}

\noindent we can write, e.g., $F(k+1)$ again in terms of $n$, $k$, $\binom{n}{k}$ and $S_1(k)$:

\vspace*{-0.4cm}

\begin{equation}\label{Equ:A1SummandInSeq}
F(k+1)=\SH_k F(k)=\big(1-(n-2(k+1))\big(S_1(k)+\tfrac{1}{k+1}\big)\big)\tfrac{n-k}{k+1}\tbinom{n}{k}^{-1}.
\end{equation}

\vspace*{-0.1cm}

\noindent We will utilize this property, but instead of working with the summand objects $k$, $\binom{n}{k}$ and $S_1(k)$ we will represent the objects by the variables $x$, $b$, $h$, respectively; $n$ is also considered as a variable. Here we start with the rational numbers and construct the rational function field\footnote{$\ZZ$ are the integers, $\NN=\{0,1,2,\dots\}$ are the non-negative integers, and all fields (resp.\ rings) contain the rational numbers $\QQ$ as a subfield (resp.\ subring). For a set $A$ we define $A^*:=A\setminus\{0\}$.}
$\FF:=\QQ(n)(x)(b)(h)$, i.e., the field of quotients of polynomials in the variables $n,x,b,h$. In this way, \eqref{Equ:A1SummandInSeq} is represented by

\vspace*{-0.2cm}

\begin{equation}\label{Equ:A1SummandInF}
f=\big(1-(n-2(x+1))(h+\frac{1}{x+1}\big)\frac{n-x}{x+1}b^{-1}\in\FF.
\end{equation}

\noindent Finally, we model the shift operator $\SH_k$ by a field automorphism 
$\fct{\sigma}{\FF}{\FF}$.

\begin{definition}\label{Def:RingAuto}
Let $\FF$ be a field (resp.\ ring). A bijective map $\fct{\sigma}{\FF}{\FF}$ is called \textit{field} (resp.\ \textit{ring}) \textit{automorphism} if $\sigma(a\circ b)=\sigma(a)\circ \sigma(b)$ for all $a,b\in\FF$ and $\circ\in\{+,\cdot\}$.
\end{definition}

\noindent
\textit{Remark.} If $\FF$ is a ring, it follows that $\sigma(0)=0$, $\sigma(1)=1$ and $\sigma(-a)=-\sigma(a)$ for all $a\in\FF$. In addition, if $\FF$ is a field, this implies that $\sigma(1/a)=1/\sigma(a)$ for all $a\in\FF^*$.

\smallskip

\noindent Namely, looking at the shift behavior of the summand objects~\eqref{Equ:SummandShiftBehavior} the automorphism is constructed as follows. We start with the rational function field $\QQ(n)$ and define $\fct{\sigma}{\QQ(n)}{\QQ(n)}$ with $\sigma(c)=c$ for all $c\in\QQ(n)$.
Next, we extend $\sigma$ to $\QQ(n)(x)$ such that $\sigma(x)=x+1$. We note that this construction is unique:

\begin{lemma}
Let $\FF(t)$ be a rational function field, $\fct{\sigma}{\FF}{\FF}$ be a field automorphism, and $a,b\in\FF$ with $a\neq0$. Then there is exactly one way how the field automorphism is extended to $\FF(t)$ subject to the relation $\sigma(t)=a\,t+b$. Namely, for $f=\sum_{i=0}^n f_i t^i\in\FF[t]$, $\sigma(f)=\sum_{i=0}^n\sigma(f_i)(a\,t+b)^i$. And for $p,q\in\FF[t]$ with $q\neq0$, $\sigma(\frac{p}{q})=\frac{\sigma(p)}{\sigma(q)}$.
\end{lemma}

\noindent As a consequence, by iterative application we extend $\sigma$ uniquely from $\QQ(n)$ to $\QQ(n)(x)(b)(h)$ subject to the shift relations (compare~\eqref{Equ:SummandShiftBehavior})
\begin{align}\label{Equ:A1Automorph}
\sigma(x)&=x+1,&
\sigma(b)&=\frac{n-x}{x+1}b,&
\sigma(h)&=h+\frac{1}{x+1}.
\end{align}
In summary, we represent the summand $F(k+1)$ given in~\eqref{Equ:A1SummandInSeq} by~\eqref{Equ:A1SummandInF} in the rational function field  $\FF:=\QQ(n)(x)(b)(h)$ together with its field automorphism  $\fct{\sigma}{\FF}{\FF}$ subject to the shift relations~\eqref{Equ:A1Automorph}. Exactly this construction is called difference field; for a general theory see~\cite{Cohn:65,Levin:08}.

\begin{definition}
A \textit{difference field}\index{difference field}\index{difference ring} (resp.\ \textit{difference ring}) $\dfield{\FF}{\sigma}$ is a field (resp.\ ring) $\FF$ together with a field automorphism (resp.\ ring automorphism) $\fct{\sigma}{\FF}{\FF}$. Here we define the set of constants by
$\Dconst{\FF}{\sigma}:=\{c\in\FF|\sigma(c)=c\}.$
\end{definition}
\textit{Remark.} For a difference field $\dfield{\FF}{\sigma}$ the set $\Dconst{\FF}{\sigma}$ forms a subfield $\FF$ which is also called \textit{constant field}\index{constant field} of $\dfield{\FF}{\sigma}$. Since $\QQ$ is always kept invariant under $\sigma$ (this is a consequence of $\sigma(1)=1$), $\QQ$ is always contained in $\Dconst{\FF}{\sigma}$ as a subfield.

\smallskip

\noindent We continue with our concrete example. Given the difference field $\dfield{\FF}{\sigma}$ in which $F(k+1)$ is represented by~\eqref{Equ:A1SummandInF}, we search for a rational function $g\in\FF$ such that
\begin{equation}\label{Equ:TeleDF}
\sigma(g)=g+f.
\end{equation}
Namely, we activate the algorithm from Section~\ref{Sec:TeleAlg} below and calculate the solution 
\begin{equation}\label{Equ:TeleSol}
g=((x+1)h+1)b^{-1}+c,\quad\quad c\in\QQ(n)
\end{equation} 
which rephrased in terms of the summation objects gives the solution~\eqref{Equ:TeleSumSol} for~\eqref{Equ:TeleSum}.

In a nutshell, the proposed simplification tactic consists of the following steps.

\begin{enumerate}
\item Construct a difference field in which the summand objects can be rephrased.
\item Find a solution $g$ of~\eqref{Equ:TeleDF} in this difference field (or a suitable extension).  
\item Reformulate $g$ to a solution $G(k)$ of~\eqref{Equ:TeleSum} in terms of product-sum expressions.
\end{enumerate}

\noindent The algorithms in the next section deliver tools to attack this problem for the class of indefinite nested product-sum expression; for a more formal framework see~\cite{Schneider:10b}.

\begin{definition}\label{Def:IndefiniteNestedSums}
Let $\KK$ be a field and the variable $k$ be algebraically independent over $\KK$.
An expression is called \textit{indefinite nested product-sum expression w.r.t.\ $k$}
iff it can be built by $k$, a finite number of constants from $\KK$, the four operations ($+,-,\cdot,/$), and
sums and products of the type  
$\sum_{i=l}^kf(i)$ or $\prod_{i=l}^kf(i)$ where $l\in\NN$ and where $f(i)$ is an indefinite nested product-sum expression w.r.t.\ $i$ which is free of $k$ . 
In particular, we require that there is a $\lambda\in\NN$ such that for any integer $n\geq\lambda$ the expression evaluates for $k=n$ (to an element form $\KK$) without entering in any pole.
\end{definition}

\noindent In this recursive definition the sums and products can be arbitrarily composed and can arise also as polynomial expressions in the denominators. Here we restrict ourselves to those expressions that occurred in practical problem solving so far. 

\begin{definition}\label{Def:NestedHgSums}
A sequence $\langle h_u\rangle_{u\geq0}\in\KK^{\NN}$ is called \textit{hypergeometric}\index{sequence!hypergeometric}  if there are $\alpha(x)\in\KK(x)$ and $l\in\NN$ such that $h_{u+1}/h_u=\alpha(u)$ for all $u\geq l$. I.e., $h_u=c\,\prod_{i=l+1}^u\alpha(i-1)$ for all $u\geq l$ for some $c\in\KK^*$. Such a symbolic product ($u$ replaced by a variable $k$) is called \textit{hypergeometric product}\index{expression!hypergeometric product} w.r.t.\ $k$.
An expression is called \textit{nested hypergeometric sum expression}\index{expression!nested hypergeometric sum}\index{nested hypergeometric sum expression} w.r.t.\ $k$ if it is an indefinite nested product-sum expression w.r.t.\ $k$ (see Def.~\ref{Def:IndefiniteNestedSums}) such that the arising products are hypergeometric and the arising sums and products occur only as polynomial expressions in the numerators. The arising sums (with upper bound $k$) are called \textit{nested hypergeometric sums (w.r.t.\ $k$)}.
\end{definition}

\noindent E.g., the harmonic sums~\eqref{Equ:HarmonicSums} and their generalizations~\cite{Moch:02,ABS:11} fall into this class. In particular, the right hand sides of~\eqref{Id:AM1}--\eqref{Id:AM4} are covered.
These expressions evaluate exactly to the d'Alembertian sequences\index{sequence!d'Alembertian}~\cite{Abramov:94},\cite{Petkov:2013}. 

\begin{remark}\label{Remark:DFToEvSolver}
In the difference field approach also the $q$-hypergeometric and mixed case~\cite{Bauer:99} can be handled.
All what will follow generalizes to this extended setting. 
\end{remark}


Subsequently, we will derive a full algorithm that treats the three steps from above automatically for the class of nested hypergeometric sum expressions.

\section{Details of the difference field machinery}\label{Sec:DFDetails}

We will work out how the Steps 1--3 from above (covering also the more general paradigms of creative telescoping and recurrence solving) can be carried out automatically. 
As an important consequence we will obtain tools to compactify nested hypergeometric sum expressions, i.e., the occurring sums in the derived expression are algebraically independent (see also Section in~\ref{Sec:IndefiniteSummation}). 


\subsection{Step 1: From indefinite nested sums and products to \pisiSE-fields}\label{Sec:Step1}

In the previous section the construction of a difference field for a given summand in terms of indefinite nested product-sums was as follows.
We start with a constant difference field $\dfield{\KK}{\sigma}$, i.e.,  $\sigma(c)=c$ for all $c\in\KK$ or equivalently $\Dconst{\KK}{\sigma}=\KK$.
Then we adjoin step by step new variables, say $t_1,\dots,t_e$ to $\KK$ which gives the rational function field $\FF:=\KK(t_1)(t_2)\dots(t_e)$ and extend the field automorphism from $\KK$ to $\FF$ subject to the shift relations $\sigma(t_i)=a_i\,t_i$ or
$\sigma(t_i)=t_i+a_i$ for some $a_i\in\FF(t_1)\dots(t_{i-1})^*$.
Subsequently, we restrict this construction
to \pisiSE-fields; for a slightly more general but rather technical definition of Karr's $\Pi\Sigma$-fields see~\cite{Karr:81,Karr:85}.

\begin{definition}\label{Def:PiSiField}
$\dfield{\FF}{\sigma}$ as given above is called \textit{\pisiSE-field over $\KK$}\index{difference field!\pisiSE-field} if $\Dconst{\FF}{\sigma}=\KK$. The adjoined elements $(t_1,\dots,t_e)$ are also called generators of the \pisiSE-field.
\end{definition}

\noindent E.g., our difference field $\dfield{\QQ(n)(x)(b)(h)}{\sigma}$ with~\eqref{Equ:A1Automorph} is a
\pisiSE-field over $\QQ(n)$. 
To see that the constants are just $\QQ(n)$, the following result is crucial~\cite{Karr:81,Schneider:01}.

\begin{theorem}\label{Thm:Karr}[Karr's theorem]
Let $\dfield{\FF}{\sigma}$ be a difference field, take a rational function field $\FF(t)$, and extend the automorphism $\sigma$ from $\FF$ to $\FF(t)$ subject to the relation $\sigma(t)=a\,t+f$ for some $a\in\FF^*$ and $f\in\FF$. Then the following holds.\\
1) Case $a=1$: $\Dconst{\FF(t)}{\sigma}=\Dconst{\FF}{\sigma}$ iff there is no $g\in\FF$ with $\sigma(g)=g+f$.\\
2) Case $f=0$: $\Dconst{\FF(t)}{\sigma}=\Dconst{\FF}{\sigma}$ iff there is no $g\in\FF^*$, $r>0$ with $\sigma(g)=a^r\,g$.
\end{theorem}

\begin{example}\label{Exp:ConstructPiSi}
Using Theorem~\ref{Thm:Karr} we represent the sum~\eqref{Equ:A1} in a \pisiSE-field parsing the occurring objects in the following order: $\;\;\stackrel{(0)}{\to}n\stackrel{(1)}{\to} k \stackrel{(2)}{\to} \binom{n}{k} \stackrel{(3)}{\to} S_1(k) \stackrel{(4)}{\to} A_{-1}(a)$.\\ 
(0) We start with $\dfield{\QQ(n)}{\sigma}$ setting $\sigma(c)=c$ for all $c\in\QQ(n)$.\\
(1) Then we construct the difference field $\dfield{\QQ(n)(x)}{\sigma}$ subject to the shift relation $\sigma(x)=x+1$. Since there is no $g\in\QQ(n)$ such that $\sigma(g)=g+1$, it follows by Karr's Theorem that $\Dconst{\QQ(n)(x)}{\sigma}=\QQ(n)$, i.e., $\dfield{\QQ(n)(x)}{\sigma}$ is a \pisiSE-field over $\QQ(n)$.\\
(2) One can check by an algorithm of Karr~\cite{Karr:81} that there is no $r>0$ and $g\in\QQ(n)(x)^*$ such that $\sigma(g)=\left(\frac{n-x}{x+1}\right)^rg$. Thus for our difference field $\dfield{\QQ(n)(x)(b)}{\sigma}$ with $\sigma(b)=\frac{n-x}{x+1}b$ we have that $\Dconst{\QQ(n)(x)(b)}{\sigma}=\Dconst{\QQ(n)(x)}{\sigma}=\QQ(n)$ by Theorem~\ref{Thm:Karr}, i.e., $\dfield{\QQ(n)(x)(b)}{\sigma}$  is a \pisiSE-field over $\QQ(n)$.\\
(3) Next, we extend the \pisiSE-field $\dfield{\QQ(n)(x)(b)}{\sigma}$ to $\dfield{\QQ(n)(x)(b)(h)}{\sigma}$
subject to the shift relation $\sigma(h)=h+\frac{1}{x+1}$. There is no $g\in\QQ(n)(x)(b)$ with $\sigma(g)=g+\frac{1}{x+1}$; this can be checked by the algorithm given in Section~\ref{Sec:TeleAlg} below. Thus  the constants remain unchanged by Theorem~\ref{Thm:Karr}, and $\dfield{\QQ(n)(x)(b)(h)}{\sigma}$ is a \pisiSE-field over $\QQ(n)$.\\
(4) Given $f$ in~\eqref{Equ:A1SummandInF}, that represents $F(k+1)$ in~\eqref{Equ:A1}, we find~\eqref{Equ:TeleSol} such that $\sigma(g)=g+f$. In other words, $g$ reflects the shift behavior of $A_{-1}(k)=\sum_{i=1}^kF(i)$ with $\SH_k A_{-1}(k)=A_{-1}(k)+F(k+1)$. Reformulating $g$ in terms of sums and products yields~\eqref{Equ:TeleSumSol} and choosing $c=0$ delivers the identity~\eqref{Equ:A1Indef}. In other words, $g$ (for $c=0$) can be identified with the sum $A_{-1}(k)$. This construction will be done more precise in Subsection~\ref{Sec:Step3}; in particular, we refer to Remark~\ref{Remark:SimultaneousConstr}.  
\end{example}

\noindent In general, Theorem~\ref{Thm:Karr} yields the following telescoping tactic to represent a given indefinite nested product-sum expression (see Definition~\ref{Def:IndefiniteNestedSums}) in terms of a \pisiSE-field. One starts with the constant field $\dfield{\KK}{\sigma}$ with $\sigma(c)=c$ for all $c\in\KK$. Then one parses all the summation objects. Suppose one treats in the next step a sum of the form $\sum_{i=1}^kF(i)$ where one can
express $F(k)$ in the so far constructed \pisiSE-field $\dfield{\FF}{\sigma}$, say $F(k+1)$ can be rephrased by $f\in\FF$. Then there are two cases: one finds a $g\in\FF$ such $\sigma(g)=g+f$ and one can model the sum $\sum_{i=1}^kF(i)$ with its shift behavior 
\begin{equation}\label{Equ:SumShiftB}
\SH_{k} \sum_{i=1}^kF(i)=\sum_{i=1}^kF(i)+F(k+1)
\end{equation}
by $g+c$ (for some properly chosen $c\in\KK$). If this fails, one can adjoin a new variable, say $t$, to $\FF$ and extends the automorphism to $\fct{\sigma}{\FF(t)}{\FF(t)}$ subject to the shift relation $\sigma(t)=t+f$. By Theorem~\ref{Thm:Karr} the constants remain unchanged, i.e., $\dfield{\FF(t)}{\sigma}$ is a \pisiSE-field over $\KK$, and $t\in\FF(t)$  models accordingly the shift behavior~\eqref{Equ:SumShiftB} of our sum. The product case can be treated similarly; see also Problem~RP on page~\pageref{ProblemRP}.

\subsection{Step 2: Solving the telescoping problem in a given \pisiSE-field}\label{Sec:TeleAlg}

Karr's algorithm~\cite{Karr:81} solves the telescoping problem within a fixed \pisiSE-field exploiting its recursive nature: it tries to solve the problem for the top most generator and reduces the problem to the subfield (i.e., without the top generator). This reduction is possible by solving the following more general problem.

\smallskip

\noindent\MyFrame{\noindent\textbf{Problem FPLDE: First-order Parameterized Linear Difference Equations.}\index{difference equation!parameterized first-order}\\
\textit{Given} a \pisiSE-field $\dfield{\FF}{\sigma}$ over $\KK$, $\alpha_0,\alpha_1\in\FF^*$ and $f_0,\dots,f_d\in\FF$.\\
\textit{Find} all\footnote{The solution set $V=\{(c_0,\dots,c_d,g)\in\KK^{d+1}\times\FF|\alpha_1\sigma(g)+\alpha_0g=\sum_{i=0}^dc_if_i\}$ forms a $\KK$-vector space of dimension $\leq d+2$ and the algorithm calculates an explicit basis of $V$.} $c_0,\dots,c_d\in\KK$ and $g\in\FF$ such that 
$\alpha_1\sigma(g)+\alpha_0g=c_0f_0+\dots+c_df_d$.}\\[0.08cm]
\textit{Remark.} Problem~FPLDE contains not only the summation paradigm of telescoping, but also of creative (resp.\ parameterized) telescoping~\eqref{Equ:ParaTele} for a fixed \pisiSE-field.

\smallskip

\noindent Subsequently, we sketch a simplified version of Karr's algorithm applied to our concrete problem:
Given the \pisiSE-field
$\dfield{\FF}{\sigma}$ with $\FF=\QQ(n)(x)(b)(h)$ and the shift relations~\eqref{Equ:A1Automorph} and given the summand~\eqref{Equ:A1SummandInF}, calculate (if possible) $g\in\FF$ such that $\sigma(g)-g=f$ holds. 
The algorithm is recursive: it treats the top most variable $h$ and needs to solve FPLDEs in the smaller \pisiSE-field $\dfield{\HF}{\sigma}$ with $\HF=\QQ(n)(x)(b)$.

\smallskip

\noindent \textit{Denominator bounding:} Calculate a polynomial $q\in\HF[h]^*$ such that for any $g\in\HF(h)$ with~\eqref{Equ:TeleDF} we have that $g\,q\in\HF[h]$, i.e., $q$ contains the denominators of all the solutions as a factor. For a general \pisiSE-field and $f$ such a universal denominator $q$ can be calculated; see~\cite{Karr:81,Bron:00,Schneider:04b}. Then given such a $q$, it suffices to search for a polynomial $p\in\HF[h]$ such that the first order difference equation 
\begin{equation}\label{Equ:DenProb}
\frac{1}{\sigma(q)}\sigma(p)-\frac{1}{q}p=f
\end{equation}
holds (which is covered by Problem~FPLDE). In our concrete example the algorithm outputs that we can choose $q=1$, i.e., we have to search for a $p\in\HF[h]$ such that 
$\sigma(p)-p=f$ holds.

\smallskip

\noindent \textit{Degree bounding:} Calculate $b$ such that for any $p\in\HF[h]$ with~\eqref{Equ:DenProb} we have that $\deg(p)\leq b$. For a general \pisiSE-field and $f$ such a $b$ can be calculated; see~\cite{Karr:81,Schneider:05b}. In our concrete example we get $b=2$. Hence, any solution $p\in\HF[h]$ of~$\sigma(p)-p=f$ is of the form $p=p_2h^2+p_1h+p_0$ and it remains to determine $p_2,p_1,p_0\in\HF$.

\smallskip

\noindent\textit{Degree reduction:} By coefficient comparison of $h^2$ in
\begin{equation}\label{Equ:TeleAnsatz}
\sigma(p_2h^2+p_1h^1+p_0)-(p_2h^2+p_1h^1+p_0)=f
\end{equation}
we obtain the constraint $\sigma(p_2)-p_2=0$ on $p_2$. Since $\dfield{\HF}{\sigma}$ is a \pisiSE-field, $p_2\in\QQ(n)$. Hence we can choose $p_2=d$ where $d\in\QQ(n)$ is (at this point) free to choose.
Now we move $p_2 h^2=d\,h^2$ in~\eqref{Equ:TeleAnsatz} to the other side and get the equation
\begin{equation}\label{Equ:PolySolDeg1}
\sigma(p_1h^1+p_0)-(p_1h^1+p_0)=f-d\tfrac{2 h (x+1)+1}{(x+1)^2}.
\end{equation}
Note that we accomplished a simplification: the degree of $h$ in the difference equation is reduced (with the price to introduce the constant $d$). 
Now we repeat this degree reduction process. By coefficient comparison of $h^1$ in~\eqref{Equ:PolySolDeg1} we get the constraint
$\sigma(p_1)-p_1=\frac{(x+1) (2
   x-n+2)}{b
   (n-x)}+d\tfrac{-2}{x+1}$
on $p_1$. Again we succeeded in a  reduction: we have to solve Problem~FPLDE in $\HF$. Applying the sketched method recursively, gives the generic solution $d=0$ and 
$p_1=\frac{x+1}{b}+e$ with $e\in\QQ(n)$. Plugging this solution into~\eqref{Equ:PolySolDeg1} and bringing $\sigma(p_1\,h)-p_1\,h$ to the right hand side reduce the problem to
$\sigma(p_0)-p_0=\frac{-2 x+n-1}{b (x-n)}+e\frac{-1}{x+1};$
note that we decreased the degree of $h$ from $1$ to $0$, i.e., we have to solve again Problem~FPLDE in $\HF$.
Recursive application of the algorithm calculates the generic solution $e=0$ and $p_0=b^{-1}+c$ with $c\in\QQ(n)$. Putting everything together gives the solution~\eqref{Equ:TeleSol}.

The technical details of the sketched algorithm for Problem~FPLDE can be found in~\cite{Schneider:04a,Bron:00}. More generally, this algorithm can be extended to a method from the first-order case to the $m$th-order case ($m\in\NN$) as described in~\cite{Schneider:05a}. Furthermore, taking results from~\cite{ABPS:13} we obtain a full algorithm that solves the following 

\medskip

\noindent\MyFrame{\noindent\textbf{Key problem PLDE: Parameterized Linear Difference Equations.}\index{difference equation!parameterized higher-order}\\
\textit{Given} a \pisiSE-field $\dfield{\FF}{\sigma}$ over $\KK$, $\alpha_0,\dots,\alpha_m\in\FF$ (not all zero) and $f_0,\dots,f_d\in\FF$.\\
\textit{Find} all\footnote{The solution set $V=\{(c_0,\dots,c_d,g)\in\KK^{d+1}\times\FF\;|\;\eqref{Equ:PLDE}\text{ holds}\}$ forms a $\KK$-vector space of dimension $\leq m+d+1$ and the algorithm calculates an explicit basis of $V$.} $c_0,\dots,c_d\in\KK$ and $g\in\FF$ such that 
\begin{equation}\label{Equ:PLDE}
\alpha_m\sigma^m(g)+\dots+\alpha_0g=c_0f_0+\dots+c_df_d.
\end{equation}}\\[0.2cm]
\textit{Remark.} Problem~PLDE covers telescoping (see~\eqref{Equ:TeleDF}), creative telescoping (see~\eqref{Equ:ParaTele}) and recurrence solving (see~\eqref{Equ:LinDiffEqu}) for a given \pisiSE-field. In particular, it is a crucial building block for the enhanced summation paradigms given below. Furthermore, it allows to deal with holonomic sequences\index{summation!holonomic sequences} in the setting of difference fields~\cite{Schneider:05d,Schneider:07a}.

\subsection{Restriction to polynomial \pisiSE-fields}\label{Sec:Step1R}

We described how the summation objects can be rephrased in a \pisiSE-field (Step 1) and how the telescoping problem, and more generally Problems~FPLDE and PLDE can be solved there (Step 2). 
Subsequently, we restrict to polynomial \pisiSE-fields. This will allow us to reformulate the found result completely automatically in terms of the given summation objects (Step 3) in Subsection~\ref{Sec:Step3}.

\begin{definition}\label{Def:PolyPiSi} A \pisiSE-field\index{difference field!\pisiSE-field!polynomial}  $\dfield{\FF}{\sigma}$ over $\KK$ with $\FF=\KK(x)(p_1)\dots(p_r)(s_1)\dots(s_e)$ is called \textit{polynomial} if $\sigma(x)=x+1$, 

\vspace{-0.2cm}

\begin{itemize}
\item $\sigma(p_i)=a_i\,p_i$ with $a_i\in\KK(x)^*$ for all $1\leq i\leq r$, and
\item  $\sigma(s_i)=s_i+f_i,$ with\footnote{$\KK(x)[p_1,p_1^{-1},\dots,p_{r},p_{r}^{-1}]$ stands for the polynomial Laurent ring in the variables $p_1,\dots,p_r$, i.e., an element is of the form $\sum_{(i_1,\dots,i_r)\in S}f_{(i_1,\dots,i_r)}p_1^{i_1}\dots p_r^{i_r}$ where $f_{(i_1,\dots,i_r)}\in\KK(x)$ and $S\subseteq\set Z^r$ is finite.}
$f_i\in\KK(x)[p_1,p_1^{-1},\dots,p_{r},p_{r}^{-1}][s_1,\dots,s_{i-1}]$ for all $1\leq i\leq e$.
\end{itemize}
\end{definition}

\noindent Related to Remark~\ref{Remark:DFToEvSolver} we note that in~\cite{Schneider:10c} a more general definition is used that covers also the $q$-hypergeometric and mixed case~\cite{Bauer:99}.
All what will follow generalizes to this general setting. 
Let $\dfield{\FF}{\sigma}$ be a polynomial \pisiSE-field over $\KK$ as in Definition~\ref{Def:PolyPiSi} and define
the ring 
\begin{equation}\label{Equ:PSPolyRing}
R=\KK(x)[p_1,p_1^{-1},\dots,p_{r},p_{r}^{-1}][s_1,\dots,s_{e}].
\end{equation} 
Note that for all $g\in R$ and $k\in\ZZ$ we have that $\sigma^{k}(g)\in R$. Thus restricting $\sigma$ to $R$ gives a ring automorphism (see Definition~\ref{Def:RingAuto}). Therefore $\dfield{R}{\sigma}$ is a difference ring and the set of constants is the field $\KK$.

\begin{example}[See Ex.~\ref{Exp:ConstructPiSi}]\label{Exp:PiSiRing}
$\dfield{\QQ(n)(x)(b)(h)}{\sigma}$ is a polynomial \pisiSE-field over $\QQ(n)$. In particular, we get the difference ring $\dfield{R}{\sigma}$ with constant field $\QQ(n)$ for the polynomial (Laurent) ring 
\begin{equation}\label{Equ:ExpR}
R=\QQ(n)(x)[b,b^{-1}][h].
\end{equation}
\end{example}

\noindent We highlight that polynomial \pisiSE-fields cover (up to the alternating sign) all \textit{nested hypergeometric sums}\index{nested hypergeometric sum} (see Definition~\ref{Def:NestedHgSums}). This can be seen as follows.

\smallskip

\noindent$\bullet$ \textit{Hypergeometric sequences.}\index{hypergeometric sequence} 
Consider, e.g., the hypergeometric products
\begin{equation}\label{Equ:HypSeqSet}
H_1(k)=\prod_{i=l_1}^k\alpha_1(i-1),\;\dots\;,H_v(k)=\prod_{i=l_v}^k\alpha_v(i-1)\quad\quad\text{with } \alpha_j(x)\in\KK(x),
\end{equation}
($H_j(k)\neq0$ for all $k\geq0$) where $\KK=\QQ$ or $\KK=\QQ(n_1,\dots,n_u)$ is a rational function field. Then there is an algorithm~\cite[Theorem~6.10]{Schneider:05c} based on Theorem~\ref{Thm:Karr} that solves

\smallskip

\noindent\MyFrame{\noindent \textbf{Problem RP: Represent Products.}\label{ProblemRP}\\
\textit{Given} a \pisiSE-field $\dfield{\KK(x)}{\sigma}$  over $\KK$ with $\sigma(x)=x+1$; $\alpha_1,\dots,\alpha_v\in\KK(x)^*$.\\
\textit{Find} a \pisiSE-field $\dfield{\FF}{\sigma}$ over $\KK$ with $\FF=\KK(x)(p_1)\dots(p_r)$ and $\sigma(p_i)/p_i\in\KK(x)$ for $1\leq i\leq r$  together with  $g_j\in \KK(x)[p_1,p_1^{-1},\dots,p_{r},p_{r}^{-1}]^*$ and $b_j\in\{-1,1\}$ for $1\leq j\leq v$ such that
$\sigma(g_j)=b_j\alpha_j\,g_j.$}

\medskip

\noindent Namely, given $\alpha_1,\dots,\alpha_v\in\KK(x)$, let $\dfield{\FF}{\sigma}$ with $R:=\KK(x)[p_1,p_1^{-1},\dots,p_{r},p_{r}^{-1}]$ together with $g_j\in R$ and $b_j\in\{-1,1\}$ for $1\leq j\leq v$ be the output of Problem~RP.\\
If $b_j=1$ for all $1\leq j\leq v$, the products $H_j(k)$ in~\eqref{Equ:HypSeqSet} can be expressed with $c_j\,g_j$ for appropriate $c_j\in\KK^*$ in the polynomial \pisiSE-field $\dfield{\FF}{\sigma}$.\\ Otherwise, construct the difference ring $\dfield{\FF[m]}{\sigma}$ with $\sigma(m)=-m$, $m^2=1$ and $\Dconst{\FF[m]}{\sigma}=\KK$; see~\cite{Schneider:01}. Here $m$ models $(-1)^k$ with $\SH_k(-1)^k=-(-1)^k$.
Then the $H_j(k)$ in~\eqref{Equ:HypSeqSet} are rephrased with $c_j\,m^{(1-b_j)/2}\,g_j$ for appropriate $c_j\in\KK^*$.

\smallskip

\noindent$\bullet$ \textit{Indefinite nested sums.} Take an expression in terms of nested hypergeometric sums, i.e., the sums do not occur in a denominator. Moreover, suppose that all the arising hypergeometric products can be expressed in a polynomial \pisiSE-field. Thus it remains to deal only with summation signs and to extend the given polynomial \pisiSE-field using Theorem~\ref{Thm:Karr}.1.
Suppose that during this construction it was so far possible to obtain a polynomial \pisiSE-field, say it is of the form $\dfield{\FF}{\sigma}$ as given in Definition~\ref{Def:PolyPiSi} with the difference ring $\dfield{R}{\sigma}$ with $R$ as in~\eqref{Equ:PSPolyRing},  and let $f\in R$ be the summand of the next sum under consideration. Then there are two cases. If we fail to find a $g\in\FF$ such that $\sigma(g)=g+f$ then we can construct the \pisiSE-field $\dfield{\FF(t)}{\sigma}$ with $\sigma(t)=t+f$ by Theorem~\ref{Thm:Karr}. In particular, this \pisiSE-field is again polynomial. Otherwise, if we find such a $g\in\FF$ with $\sigma(g)-g=f\in R$, we can apply the following result; the proof is a slight extension of the one given in~\cite[Thm.~2.7]{Schneider:10c}.

\begin{theorem}
Let $\dfield{\FF}{\sigma}$ be a polynomial \pisiSE-field over $\KK$ and consider the difference ring $(R,\sigma)$ as above. Let $g\in\FF$. If $\sigma(g)-g\in R$ then $g\in R$.
\end{theorem}

\begin{example}[Cont. Example~\ref{Exp:PiSiRing}]
For $f\in R$ with~\eqref{Equ:A1SummandInF} it follows that any solution $g\in\FF$ with $\sigma(g)=g+f$ is in $R$. Indeed, we calculated~\eqref{Equ:TeleSol}.
\end{example}

\noindent Thus we always have $g\in R$. As a consequence we can express the sum over $f$ with $g+c$ for some properly chosen $c\in\KK$ (see Remark~\ref{Remark:SimultaneousConstr}). Hence by iterative application of the above construction we never enter in the case that sums occur in the denominators. Consequently, a nested hypergeometric sum expression can be rephrased in a polynomial \pisiSE-field up to the following technical aspect.

\begin{remark}\label{Remark:AlternatingExt}
If the hypergeometric products~\eqref{Equ:HypSeqSet} cannot be expressed in a \pisiSE-field solving Problem~RP, one needs in addition the alternating sign $(-1)^k$ in the setting of difference rings; note that here we cannot work anymore with fields, since zero divisors pop up: $(1+(-1)^k)(1-(-1)^k)=0$. For simplicity, we restrict ourselves to polynomial \pisiSE-fields; the described techniques and algorithms in this article can be extended for the more technical case allowing also $(-1)^k$; see~\cite{Schneider:01,Erocal:11}.
\end{remark}

\subsection{Step 3: Evaluating elements from a \pisiSE-field to sequences}\label{Sec:Step3}

Let $\dfield{\FF}{\sigma}$ be a polynomial \pisiSE-field over $\KK$ as in Definition~\ref{Def:PolyPiSi} and define $R$ by~\eqref{Equ:PSPolyRing}. In this section we make the step precise how  elements from $R$ can be reformulated as a nested hypergeometric sum expression. I.e., how such an element $f\in R$ can be mapped via an evaluation function $\fct{\ev}{R\times\NN}{\KK}$ to a sequence $\langle \ev(f,i)\rangle_{i\geq0}$ by using an explicitly given nested hypergeometric sum expression w.r.t.\ a variable $k$. 

Before we start with a concrete example, we emphasize that this map $\ev$
should respect the ring structure $R$ and the ring automorphism $\sigma$ as follows.

\begin{definition}\label{Def:EvProperties}
A map $\fct{\ev}{R\times\NN}{\KK}$ for a difference ring $\dfield{R}{\sigma}$ with constant field $\KK$ is called \textit{evaluation function}\index{evaluation function} if the following evaluation properties hold: For all
$c\in\KK$ and all $i\geq0$ we have that $\ev(c,i)=c$, for all $f,g\in R$ there is a $\delta\geq 0$ with
\begin{align}
\forall i\geq\delta:&\;\ev(f\,g,i)=\ev(f,i)\,\ev(g,i),\label{Ev:Mult}\\
\forall i\geq\delta:&\;\ev(f+g,i)=\ev(f,i)+\ev(g,i)\label{Ev:Add};\\
\intertext{and for all $f\in R$ and $j\in\ZZ$ there is a $\delta\geq 0$ with}
\label{Ev:Shift}
\forall i\geq\delta:&\;\ev(\sigma^j(f),i)=\ev(f,i+j).
\end{align}
\end{definition}

\begin{example}[See Ex.~\ref{Exp:ConstructPiSi}]\label{Exp:EvaluationDef}
Take the polynomial \pisiSE-field $\dfield{\QQ(n)(x)(b)(h)}{\sigma}$ with \eqref{Equ:A1Automorph} and consider $\dfield{R}{\sigma}$ with~\eqref{Equ:ExpR} and constant field $\QQ(n)$. 
We construct an evaluation map $\fct{\ev}{R\times\NN}{\QQ(n)}$ as follows. For $f\in R$ and $k\in\NN$, $\ev(f,k)$ is calculated by evaluating an explicitly given nested hypergeometric sum expression.\\ 
The construction is performed iteratively following the tower of extensions in $R$.\\
(1) We define $\fct{\ev}{\QQ(n)(x)\times\NN}{\QQ(n)}$ as follows. For $\frac{p}{q}\in\QQ(n)(x)$ with $p,q\in\QQ(n)[x]$ and $\gcd(p,q)=1$,
\begin{equation}\label{Equ:EvRat}
\ev(\tfrac{p}{q},k)=
\begin{cases}
\frac{p(k)}{q(k)}&\text{if $q(k)\neq0$}\\
0&\text{if $q(k)=0$ \quad(pole case)};
\end{cases}
\end{equation}
here $p(k),q(k)$ with $k\in\NN$ denotes the evaluation of the polynomials at $x=k$. Note that the properties in Definition~\ref{Def:EvProperties} are satisfied for $\delta$ chosen sufficiently large: that is the case when one does not enter in the pole case in~\eqref{Equ:EvRat} for elements $f,g\in R$ as given in Definition~\ref{Def:EvProperties}; see also Example~\ref{Exp:LFunction}.\\
(2) Next, we extend $\ev$ from $\QQ(n)(x)$ to $\QQ(n)(x)[b,b^{-1}]$. 
We set

\vspace*{-0.1cm}

\begin{equation}\label{Equ:BinomEv}
\ev(b,k)=c_1\prod_{i=l_1}^k\frac{n+1-i}{i},\quad\quad l_1\in\set N,c_1\in\QQ(n)^*
\end{equation}

\vspace*{-0.1cm}

\noindent and prolong the ring structure as follows: for $f=\sum_{j=u}^d f_jb^j\in\QQ(n)(x)[b,b^{-1}]$ with $f_j\in\QQ(n)(x)$ and $u,d\in\set Z$ we define 
$\ev(f,k)=\sum_{j=u}^d \ev(f_j,k)\ev(b,k)^{j};$
this implies that~\eqref{Ev:Mult} and~\eqref{Ev:Add} hold for some $\delta\in\NN$ sufficiently large; see Example~\ref{Exp:LFunction}. Note that for any choice of $l_1\in\set N$ and $c\in\QQ(n)^*$ also~\eqref{Ev:Shift} is valid. Since we want to model $\binom{n}{k}=\prod_{i=1}^k\frac{n+1-i}{i}$, a natural choice is $l_1=1$, $c_1=1$.\\
(3) Finally, we set

\vspace*{-0.3cm}

\begin{equation}\label{Equ:S1Ev}
\ev(h,k)=\sum_{i=\lambda_1}^k\frac{1}{i}+d_1,\quad\quad\lambda_1\in\NN,d_1\in\QQ(n).
\end{equation}
Again properties~\eqref{Ev:Mult} and~\eqref{Ev:Add} hold for some $\delta\in\NN$ (see Example~\ref{Exp:LFunction}) if we extend $\ev$ as follows: for $f=\sum_{j=0}^df_jh^j\in R$ with $f_j\in\QQ(n)(x)[b,b^{-1}]$ we set 
$\ev(f,k)=\sum_{j=0}^d\ev(f_j,k)\ev(h,k)^j.$
In addition, property~\eqref{Ev:Shift} holds for any choice of $\lambda_1\in\NN$ and $d_1\in\QQ(n)$. Since we want to model $S_1(k)$, we take, e.g., $\lambda_1=1$ and $d_1=0$.\\
For instance, for $f\in R$ as in~\eqref{Equ:A1SummandInF} the evaluation is given by the nested hypergeometric sum expression (cf.~\eqref{Equ:A1SummandInSeq})

\vspace*{-0.2cm}

\begin{equation}\label{Equ:EvSummandForExp}
\ev(f,k)=\Big(1-(n-2(k+1))\Big(S_1(k)+\frac{1}{k+1}\Big)\Big)\frac{n-k}{k+1}\binom{n}{k}^{-1};
\end{equation}
the usage of $S_1(k)$, $\binom{n}{k}$ is just pretty printing and stands for~\eqref{Equ:BinomEv}, \eqref{Equ:S1Ev}, respectively.
\end{example}

Besides the function $\ev$ we aim at the calculation of the bounds $\delta$ in Definition~\ref{Def:EvProperties}. 

\begin{example}\label{Exp:LFunction}
For the evaluation function
$\fct{\ev}{R\times\NN}{\QQ(n)}$ from Example~\ref{Exp:EvaluationDef} the bounds can be extracted by the computable map $\fct{\beta}{R}{\NN}$ defined as follows.
For $f\in R$ let $d\in\NN$ be minimal such that for all $k\geq d$ the calculation of $\ev(f,k)$ does not enter in the pole case in~\eqref{Equ:EvRat}. More precisely,
write $f$ in the form
$f=\sum_{i\in\ZZ,j\in\NN} f_{i,j}b^ih^j$
with finitely many $f_{i,j}\in\QQ(n)(x)$ being non-zero, and choose $d\in\NN$ such that for all $i,j$ and all $k\geq d$  the denominator of $f_{i,j}$ has no pole at $x=k$. This $d$ can be calculated explicitly for any $f\in R$ and defines the function $\beta$ with  $\beta(f):=d$. Now let $f,g\in R$.
Then for $\delta:=\max(\beta(f),\beta(g))$ we have that~\eqref{Ev:Mult} and~\eqref{Ev:Add}. In addition, for all $j\in\ZZ$ choose  $\delta:=\beta(f)+\max(0,-j)$ and we get~\eqref{Ev:Shift}.
\end{example}

\noindent This example motivates the following definition~\cite{Schneider:10c}.

\begin{definition}\label{Def:LFunction}
Let $\dfield{R}{\sigma}$ be a difference ring with constant field $\KK$ and consider an evaluation function $\fct{\ev}{R\times\NN}{\KK}$. $\fct{\beta}{R}{\NN}$ is called \textit{bounding function}\index{bounding function} of $\ev$ if for all $f,g\in R$ we can take $\delta:=\max(\beta(f),\beta(g))$ such that~\eqref{Ev:Mult} and~\eqref{Ev:Add} holds, and for all $f\in R$ and $j\in\set Z$ we can take $\delta:=\beta(f)+\max(0,-j)$ such that~\eqref{Ev:Shift} holds.
\end{definition}

The concrete construction above carries over to the general case.
Let $\dfield{\FF}{\sigma}$ be a polynomial \pisiSE-field over $\KK$ as given in Definition~\ref{Def:PolyPiSi} and define $R$ by~\eqref{Equ:PSPolyRing}. Then one obtains an evaluation map $\fct{\ev}{R\times\NN}{\KK}$ in terms of explicitly given nested hypergeometric sum expressions as follows.
We start with $\fct{\ev}{\KK(x)\times\NN}{\KK}$ defined by~\eqref{Equ:EvRat}
for $\frac{p}{q}\in\KK(x)$ with $p,q\in\KK[x]$ and $\gcd(p,q)=1$. Next we define how the map acts on the $p_i$:
\begin{equation}\label{Equ:ProductEv}
\ev(p_i,k)=c_i\prod_{j=l_i}^k a_i(j-1),\quad (1\leq i\leq r);
\end{equation}
here we are free to choose $c_i\in\KK^*$, and $l_i$ is chosen such that the numerator and denominator of $a_i$ evaluated at $j$ is non-zero for all $j\geq l_i$. 
Then this map is extended to $\fct{\ev}{\bar{R}_0\times\NN}{\KK}$ with $\bar{R}_0:=\KK(x)[p_1,p_1^{-1},\dots,p_r,p_r^{-1}]$ as follows. For $f=\sum_{(i_1,\dots,i_r)\in\ZZ^r}f_{(i_1,\dots,i_r)}p_1^{i_1}\dots p_r^{i_r}\in\bar{R}_0$ with $f_{(i_1,\dots,i_r)}\in\KK(x)$ we set
$$\ev(f,k)=\sum_{(i_1,\dots,i_r)\in\ZZ^r}\ev(f_{(i_1,\dots,i_r)},k)\ev(p_1,k)^{i_1}\dots \ev(p_r,k)^{i_r}.$$
Finally, we extend iteratively this map from $\bar{R}_0$ to $R:=\bar{R}_e$. Suppose that we are given already the map for $\bar{R}_i=\bar{R}[s_i,\dots,s_{i-1}]$ with $1\leq i<e$. Then we define
\begin{equation}\label{Equ:SumEv}
\ev(s_i,k)=\sum_{j=\lambda_i}^k\ev(f_i,j-1)+d_i
\end{equation}
where $d_i\in\KK$ can be arbitrarily chosen, and $\lambda_i\in\NN$ is sufficiently large in the following sense: it is larger than the lower bounds of the arising sums and products of the explicitly given nested hypergeometric sum expression for $\ev(f_i,j-1)$ and such that during the evaluation one never enters in the pole case in~\eqref{Equ:EvRat}. 
In a nutshell, the underlying expression for $\ev(s_i,k)$ (for symbolic $k$) can be written as an indefinite nested sum without entering poles that are captured via~\eqref{Equ:EvRat}. Finally, we extend this construction to $\bar{R}_{i-1}[s_i]$: for $f=\sum_{j=0}^vf_j s_i^j\in\bar{R}_{i-1}[s_i]$ with $f_j\in\bar{R}_{i-1}$ we define 
$\ev(f,k)=\sum_{j=0}^v\ev(f_j,k)\ev(s_i,k)^j.$
To this end, by iteration on $i$ ($1\leq i\leq e$) we obtain $\fct{\ev}{R\times\NN}{\KK}$ with $R:=\bar{R}_e$ that satisfies the properties in Definition~\ref{Def:EvProperties} and which is explicitly given in terms of nested hypergeometric sum expressions.

\smallskip

\noindent\textit{Remark.} Note that the products in~\eqref{Equ:ProductEv} and sums in~\eqref{Equ:SumEv} are just nested hypergeometric sum expressions w.r.t.\ $k$; see Definition~\ref{Def:NestedHgSums}.

\smallskip

\noindent Moreover, we can define explicitly a bounding function $\fct{\beta}{R}{\NN}$ that produces the required bounds $\delta$ in Definition~\ref{Def:EvProperties} following the construction in Example~\ref{Exp:LFunction}. In short, consider $f\in R$ as a polynomial in the variables $p_i,s_i$ and take all its  coefficients from $\KK(x)$. Then $\beta(f)$ is the minimal value $d\in\NN$ such that the evaluation of the coefficients does not enter in the pole case of~\eqref{Equ:EvRat}; note that the positive integer roots of the denominators can be detected if $\KK$ is computable (in particular, if one can factorize polynomials over $\KK$). Summarizing, we get Lemma~\ref{Lemma:ContructEv}; cf.~\cite{Schneider:10c}. We can summarize this construction as follows.

\begin{lemma}\label{Lemma:ContructEv}
Let $\dfield{\FF}{\sigma}$ be a polynomial \pisiSE-field over $\KK$ with $R$ as in~\eqref{Equ:PSPolyRing}. Then $\fct{\ev}{R\times\NN}{\KK}$ defined above in terms of nested hypergeometric sum expressions is an evaluation function, and $\fct{\beta}{R}{\NN}$ given above is a corresponding bounding function of $\ev$. If $\KK$ is computable, such functions can be calculated explicitly. 
\end{lemma}

\begin{remark}\label{Remark:PLDEGen} 
\textit{Solving Problem PLDE.} Take a polynomial \pisiSE-field $\dfield{\FF}{\sigma}$ as in Definition~\ref{Def:PolyPiSi} and define $R$ by~\eqref{Equ:PSPolyRing}. Let $\fct{\ev}{R\times\NN}{\KK}$ be an evaluation function with a bounding function $\fct{\beta}{R}{\NN}$ as constructed above by means of hypergeometric sum expressions. Let $\alpha_i,f_i\in R$ and take $g\in R$ such that equation~\eqref{Equ:PLDE}\index{difference equation!parameterized}
holds. 
Now calculate  $\delta:=\max(\beta(\alpha_0),\dots,\beta(\alpha_m),\beta(f_0),\dots,\beta(f_d),\beta(g))$.
Then for any $k\geq\delta$,
\begin{equation}\label{Equ:PLDESeq}
\ev(\alpha_m,k)\ev(g,k+m)+\dots+\ev(\alpha_0,k)\ev(g,k)=c_0\ev(f_0,k)+\dots+c_d\ev(f_d,k).
\end{equation}
Hence a solution $g\in R$ of~\eqref{Equ:PLDE} produces a solution of~\eqref{Equ:PLDESeq} in terms of nested hypergeometric sum expressions. In particular, a lower bound $\delta$ for its validity can be computed (if $\KK$ is computable).
We emphasize that this property is crucial for the automatic execution of the summation paradigms given in Section~\ref{Sec:SumTechniques} below.
\end{remark}

\begin{remark}\label{Remark:SimultaneousConstr} \textit{Simultaneous construction of a \pisiSE-field and its evaluation function.} 
In order to model the summation problem accordingly (see, e.g., Problem~EAR on page~\pageref{ProblemEAR}) the construction of the \pisiSE-field (Step 1) and the evaluation function with its bounding function should be performed simultaneously. Here the choice of the lower bounds and constants in~\eqref{Equ:ProductEv} and~\eqref{Equ:SumEv} are adjusted such that the evaluation of the introduced products and sums agrees with the objects of the input expression; for a typical execution see Example~\ref{Exp:EvaluationDef}.
In particular the evaluation function is crucial if a sum can be represented in the already given \pisiSE-field by telescoping. 
This is, e.g., the case in (5) of Example~\ref{Exp:ConstructPiSi}. We succeeded in representing the summand $F(k)$ in~\eqref{Equ:A1SummandInSeq} by the element $f\in R$ as in~\eqref{Equ:A1SummandInF}. Namely, using the evaluation function from Example~\ref{Exp:EvaluationDef} we have~\eqref{Equ:EvSummandForExp} for all $k\geq0$.
Then we calculate the solution~\eqref{Equ:TeleSol} of $\sigma(g)=g+f$. In particular, we obtain
$\ev(g,k)=((k+1)S_1(k)+1)\binom{n}{k}^{-1}+c$ with $c\in\QQ(n)$
such that for all $k\geq0$ we have that
$\ev(g,k+1)=\ev(g,k)+\ev(f,k+1)$;
see Remark~\ref{Remark:PLDEGen}. Now we follow the same arguments as in the beginning of Section~\ref{Sec:BasicMechanism}: $A_{-1}(a)$ and $\ev(g,a)$ satisfy the same recurrence $A(a+1)=A(a)+\ev(f,a+1)$ for all $a\geq0$ and thus $\ev(g,a)=A_{-1}(a)$ when choosing $c=0$. In this way, we represent $A_{-1}(a)$ precisely with $g$ ($c=0$) in the given \pisiSE-field and its evaluation function.
\end{remark}

\subsection{Crucial property: algebraic independence of sequences}

Take the elements from a polynomial \pisiSE-field, rephrase them as nested hypergeometric sum expressions, and evaluate the derived objects to sequences. The main result of this subsection is that the sequences of the generators of the \pisiSE-field are algebraically independent over each other.
To see this, let $\dfield{\FF}{\sigma}$ be a polynomial \pisiSE-field over $\KK$ with 
$\FF=\KK(x)(p_1)\dots(p_r)(s_1)\dots(s_e)$ and define $R$ by~\eqref{Equ:PSPolyRing}. Moreover, take an evaluation function $\fct{\ev}{R\times\NN}{\KK}$ following the construction of the previous subsection. 
Then we can define the map $\fct{\tau}{R}{\KK^{\NN}}$ with
\begin{equation}\label{Equ:DefineTau}
\tau(f)=\langle \ev(f,k)\rangle_{k\geq0}=\langle \ev(f,0),\ev(f,1),\ev(f,2),\dots\rangle.
\end{equation}
Now we explore the connection between the difference ring $\dfield{R}{\sigma}$ and the set of sequences
$\tau(R)=\{\tau(f)|f\in R\}$.
First, we introduce the following notions.

\begin{definition}
Let $\dfield{R_1}{\sigma_1}$ and $\dfield{R_2}{\sigma_2}$ be difference rings.
\begin{itemize}
\item If $R_1$ is a subring of $R_2$ and $\sigma_1(f)=\sigma_2(f)$ for all $f\in R_1$ then $\dfield{R_1}{\sigma}$ is called \textit{sub-difference ring}\index{difference field!sub-difference field}\index{difference ring!sub-difference ring} of $\dfield{R_2}{\sigma}$.
\item A map $\fct{\tau}{R_1}{R_2}$ is called ring homomorphism if $\tau(f\,g)=\tau(f)\tau(g)$ and $\tau(f+g)=\tau(f)+\tau(g)$ for all $f,g\in R_1$. If $\tau$ is in addition injective (resp.\ bijective), $\tau$  is called \textit{ring embedding} (resp.\ \textit{ring isomorphism}). Note: if $\tau$ is an isomorphism, the rings $R_1$ and $R_2$ are the same up to renaming of the elements with $\tau$.
\item A map $\fct{\tau}{R_1}{R_2}$ is called \textit{difference ring homomorphism}\index{difference ring!homomorphism}\index{difference ring!embedding}\index{difference ring!isomorphism} (resp.\ \textit{embedding/ isomorphism}) if it is a ring homomorphism (resp.\ embedding/isomorphism) and for all $n\in\ZZ$, $f\in R_1$ we have $\tau(\sigma_1^n(f))=\sigma_2^n(\tau(f))$. Note: if $\tau$ is an isomorphism, $\dfield{R_1}{\sigma}$ and $\dfield{R_2}{\sigma}$ are the same up to renaming of the elements by $\tau$.
\end{itemize}
\end{definition}

With component-wise addition and multiplication of the elements from $\KK^{\NN}$ we obtain a commutative ring  where the multiplicative unit is $\vec{1}=\langle1,1,1,\dots\rangle$; the field $\KK$ can be
naturally embedded by mapping $k\in\KK$ to
$\vect{k}=\langle k,k,k,\dots\rangle$.

\begin{example}\label{Exp:RatSeq}
Let $\KK(x)$ be a rational function field, take the evaluation function $\fct{\ev}{\KK(x)\times\NN}{\KK(x)}$ defined by~\eqref{Equ:EvRat}, and define $\fct{\tau}{\KK(x)}{\KK^{\NN}}$ as in~\eqref{Equ:DefineTau}.
Now define the set  
$$F:=\tau(\KK(x))=\{\langle \ev(f,k)\rangle_{k\geq0}|f\in\KK(x)\}.$$
Observe that $F$ is a subring of $\KK^{\NN}$. However, it is not a field. E.g., if we multiply 
$\ev(x,k)\rangle_{k\geq 0}$ with $\ev(1/x,k)\rangle_{k\geq 0}$, we obtain
$\langle 0,1,1,1,\dots\rangle$ which is not the unit $\textbf{1}$.
But, we can turn it to a field by identifying two sequences if they agree from a certain point on. Then the inverse of $\ev(x,k)\rangle_{k\geq 0}$ is $\ev(1/x,k)\rangle_{k\geq 0}$. More generally, for $f\in\KK(x)^*$ we get $\ev(f,k)\rangle_{k\geq 0}\,\ev(1/f,k)\rangle_{k\geq 0}=\textbf{1}$. 
\end{example}

\noindent To be more precise, we follow the construction
from~\cite[Sec.~8.2]{AequalB}: We define an equivalence relation
$\sim$ on $\KK^{\NN}$ by $\langle a_n\rangle_{n\geq0}\sim
\langle b_n\rangle_{n\geq0}$ if there exists a $\delta\geq 0$ such that
$a_n=b_n$ for all $n\geq \delta$. The equivalence classes form a ring
which is denoted by $\seqR$; the elements of $\seqR$ (also called germs) will be
denoted, as above, by sequence notation. 
Finally, define the shift operator $\fct{\Shift}{\seqR}{\seqR}$ with 
\begin{equation*}
\Shift(\langle a_0,a_1,a_2,\dots\rangle)=\langle
a_1,a_2,a_3,\dots\rangle.
\end{equation*}
In this ring the shift is invertible with
$\Shift^{-1}(\langle a_1,a_2,\dots\rangle)=\langle 0,
a_1,a_2,a_3,\dots\rangle=\langle a_0,a_1,a_2,\dots\rangle.$
It is immediate that $\Shift$ is a ring automorphism and thus $\dfield{\seqR}{\Shift}$ is a difference ring. In short, we call this difference ring also \textit{ring of sequences}\index{ring of sequences}\index{sequence!difference ring}.

\begin{example}[Cont.\ Example~\ref{Exp:RatSeq}]
Consider our subring $F$ of $\seqR$. Restricting $\Shift$ to $F$ gives a bijective map and thus it is again a ring automorphism. Even more, since $F$ is a field, it is a field automorphism, and $\dfield{F}{\Shift}$ is a difference field. In particular, $\dfield{F}{\Shift}$ is a sub-difference ring of $\dfield{\seqR}{\Shift}$.
\end{example}

\noindent More generally, consider the map $\fct{\tau}{R}{\seqR}$ as in~\eqref{Equ:DefineTau}. Since $\fct{\ev}{R\times\NN}{\KK}$ has the properties as in Definition~\ref{Def:EvProperties}, it follows that for all $f,g\in R$ we have $\tau(f\,g)=\tau(f)\,\tau(g)$ and $\tau(f+g)=\tau(f)+\tau(g)$. Hence $\tau$ is a ring homomorphism. Moreover, for all $f\in R$ and all $n\in\ZZ$, $$\Shift^n(\langle\ev(f,k)\rangle_{k\geq0}\rangle)=\langle\ev(f,k+n)\rangle_{k\geq0}\rangle=\langle\ev(\sigma^n(f),k)\rangle_{k\geq0}\rangle.$$
Thus $\tau$ is a difference ring homomorphism between $\dfield{R}{\sigma}$ and $\dfield{\seqR}{\Shift}$.
Since $\tau(R)$ is a subring of $\seqR$ and $\Shift$ restricted to $\tau(R)$ is a ring automorphism, $\dfield{\tau(R)}{\Shift}$ is a difference ring, and it is a sub-difference ring of $\dfield{\seqR}{\Shift}$.

\begin{example}\label{Exp:EvExample}
Take the polynomial \pisiSE-field $\dfield{\QQ(n)(x)(b)(h)}{\sigma}$ over $\QQ(n)$ with~\eqref{Equ:A1Automorph} and~\eqref{Equ:ExpR}, and let $\fct{\ev}{R\times\NN}{\QQ(n)}$ be the evaluation function from Example~\ref{Exp:EvaluationDef}; define  $\fct{\tau}{R}{S(\QQ(n))}$ with~\eqref{Equ:DefineTau}. Then $\tau$ is a difference ring homomorphism. 
In particular, $\dfield{\tau(R)}{\Shift}$ is a difference ring and a sub-difference ring of $\dfield{S(\QQ(n))}{\Shift}$.
\end{example}

\noindent Now we can state the crucial property proven in~\cite{Schneider:10c}: our map~\eqref{Equ:DefineTau} is injective.

\begin{theorem}
Let $\dfield{\FF}{\sigma}$ be a polynomial \pisiSE-field over $\KK$, define $R$ by~\eqref{Equ:PSPolyRing}, and take an evaluation function $\fct{\ev}{R\times\NN}{\KK}$ as given in Lemma~\ref{Lemma:ContructEv}. Then $\fct{\tau}{R}{\seqR}$ with~\eqref{Equ:DefineTau} is a difference ring embedding. 
\end{theorem}

\begin{example}
$\dfield{\QQ(n)(x)}{\sigma}$ and $\dfield{\tau(\QQ(n)(x))}{\sigma}$ are isomorphic. In addition, the rings $\dfield{R}{\sigma}$ and $\dfield{\tau(R)}{\sigma}$ with $R:=\QQ(n)(x)[b,b^{-1}][h]$ are isomorphic. Thus
$\tau(R)=\tau(\QQ(n)(x))[\tau(b),\tau(b^{-1})][\tau(h)]$
is a polynomial ring and there are no algebraic relations among the sequences $\tau(b)$, $\tau(b^{-1})$, $\tau(h)$ with coefficients from $\tau(\QQ(n)(x))$.
\end{example}

\noindent In general, the difference rings $\dfield{R}{\sigma}$ and $\dfield{\tau(R)}{\Shift}$ are isomorphic: they are the same up to renaming of the elements by $\tau$. In particular, we get the polynomial ring
\begin{equation}\label{Equ:PolySequenceRing}
\tau(R)=\tau(\KK(x))[\tau(p_1),\tau(p_1^{-1}),\dots,\tau(p_{r}),\tau(p_{r}^{-1})][\tau(s_1),\dots,\tau(s_e)]
\end{equation}
with coefficients form the field $\tau(\KK(x))$. I.e., there are no 
algebraic relations among the sequences $\tau(p_i)$, $\tau(p_i^{-1})$ and $\tau(s_i)$ with coefficients from $\tau(\KK(x))$.

\section{The symbolic summation toolbox of \SigmaP}\label{Sec:SumTechniques}

In the following we will give an overview of the symbolic summation toolbox that is available in the Mathematica package \SigmaP~\cite{Schneider:07a}. Here we focus on \textit{nested hypergeometric sum expressions (w.r.t.\ $k$)}\index{nested hypergeometric sum} as given in Definition~\ref{Def:NestedHgSums}: the products are hypergeometric expressions (for more general classes see Remark~\ref{Remark:DFToEvSolver}) and the sums and products do not arise in the denominators. 

\smallskip

\noindent Concerning \textit{indefinite summation} it is shown how a nested hypergeometric sum expression can be compactified such that the arising sums are algebraically independent and such that the sums are simplified concerning certain optimality criteria. 

\smallskip

\noindent Concerning \textit{definite summation} the package \SigmaP\ provides the following toolkit.
In Section~\ref{Sec:RecurrenceFinding} it is worked out how a recurrence can be computed with creative telescoping for a definite sum over a nested hypergeometric sum expression. Moreover, in Section~\ref{Sec:RecurrenceSolving} it is elaborated how such a recurrence can be solved in terms of nested hypergeometric sum expressions which evaluate to d'Alembertian sequences. Usually the derived solutions are highly nested, and thus indefinite summation is heavily needed. Finally, given sufficiently many solutions their combination gives an alternative representation of the definite input sum. Summarizing, the following ``summation spiral'' is applied~\cite{Schneider:04c}:

\vspace*{-0.3cm}

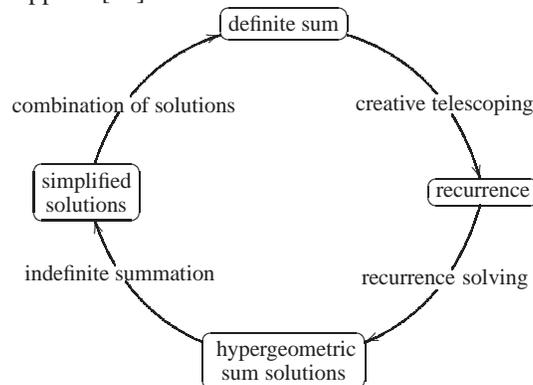
\footnotesize
$$\xymatrix@!R=1.4cm{
&*+[F-:<3pt>]{\txt{definite sum}} \ar @/^1.5pc/[rd]|>>>>>>>>>>{\txt{creative telescoping}}&\\
*+[F-:<3pt>]{\txt{simplified\\ solutions}}\ar @/^1.5pc/[ru]|>>>>>>>>>>>{\txt{combination of
solutions}} &&*+[F-:<3pt>]{\txt{recurrence}}\ar @/^1.5pc/[ld]|<<<<<<<<<<{\txt{recurrence solving}}\\
&*+[F-:<3pt>]{\txt{ 
hypergeometric\\
sum solutions}}\ar @/^1.5pc/[lu]|<<<<<<<<<{\txt{indefinite
summation}}& }$$

\normalsize

\noindent\textit{Remark.} We give details how these summation paradigms are solved in the setting of polynomial \pisiSE-fields introduced in Section~\ref{Sec:DFDetails}. These technical parts marked with~* can be ignored if one is mostly interested in applying the summation tools. 

\subsection{Simplification of nested hypergeometric sum expressions}\label{Sec:IndefiniteSummation}


All of the simplification strategies of \SigmaP\ solve the following basic problem.\\
\noindent\MyFrame{\noindent\textbf{Problem EAR: Elimination of algebraic relations\index{algebraic relation!elimination}.}\label{ProblemEAR}\\
\textit{Given}  a nested hypergeometric sum expression $F(k)$.\\
\textit{Find} a nested hypergeometric sum expression $\bar{F}(k)$ and $\lambda\in\NN$ such that $F(k)=\bar{F}(k)$ for all $k\geq\lambda$ and such that the occurring sums are algebraically independent.
}

The following solution relies on Section~\ref{Sec:DFDetails} utilizing ideas from~\cite{Karr:81,Schneider:05a,Schneider:10c}.\\
\noindent\textit{Solution$^*$.} Compute a polynomial \pisiSE-field\footnote{\label{Footnote:AlternatingSign}
As observed in Remark~\ref{Remark:AlternatingExt} one might need in addition the alternating sign to represent all hypergeometric products. The underlying solution works analogously by adapted algorithms.} $\dfield{\FF}{\sigma}$ over $\KK$ as in Definition~\ref{Equ:PSPolyRing} with $R$ defined as in~\eqref{Def:PolyPiSi} together with an evaluation function $\fct{\ev}{R\times\NN}{\KK}$ in which one obtains an explicit $f\in R$ with $\lambda\in\NN$ such that $\ev(f,k)=F(k)$ for all $k\geq\lambda$.  
Output the nested hypergeometric sum expression $\bar{F}(k)$ that encodes the evaluation $\ev(f,k)$. Concerning the algebraic independence note that the sub-difference ring~\eqref{Equ:PolySequenceRing} of the ring of sequences $\dfield{\seqR}{\Shift}$ forms a polynomial ring; here the difference ring embedding $\tau$ is defined by~\eqref{Equ:DefineTau}.
The sequences given by the objects occurring in $\bar{F}(k)$ are just the 
the generators of the polynomial ring~\eqref{Equ:PolySequenceRing}.\\[0.1cm]
\textit{Remark.} $\KK$ is the smallest field that contains the values of $F(r)$ for all $r\in\NN$ with $r\geq\lambda$. Here extra parameters are treated as variables. However, in most examples these parameters are assumed to be integer valued within a certain range. In such cases it might be necessary to adjust the summation bounds accordingly.

\smallskip

A typical instance of Problem EAR is the simplification of the sum~\eqref{Equ:A1}: Using the above technologies, see   Examples~\ref{Exp:ConstructPiSi} and~\ref{Exp:EvaluationDef} for further details,
we can reduce the sum $A_{-1}(a)$ in terms of $\binom{n}{a}$ and $S_1(a)$ and obtain the simplification~\eqref{Equ:A1Indef}.
After loading in\index{Sigma, summation package}

\begin{Cmma}
\CIn << Sigma.m\\\vspace*{-0.1cm}
\CPrint Sigma - A\; summation\; package\; by\; Carsten\;\; Schneider \copyright\ RISC\\
\end{Cmma}

\vspace*{-0.4cm}

\noindent this task can be accomplished with the function call 

\vspace*{-0.1cm}

\begin{Cmma}
\CIn SigmaReduce[\sum_{k=0}^{a}\big(1-(n-2k)S[1,k]\big)\tbinom{n}{k}^{-1},a]\\\vspace*{-0.2cm}\CMLabel{EAR1}
\COut ((a+1) S[1,a]+1)\binom{n}{a}^{-1}\\
\end{Cmma}

\vspace*{-0.1cm}
 
\noindent Note that $S[m_1,\dots,m_k,n]$ stands for the harmonic sums~\eqref{Equ:HarmonicSums}. More generally, one gets reduced representations for nested hypergeometric sum expressions such as 

\vspace*{-0.1cm}

\begin{Cmma}
\CIn\CMLabel{EAR2} SigmaReduce[\sum_{k=1}^a k^4 \tbinom{2 k}{k}^2+\frac{249}{20} \sum_{k=1}^a k^3 \
\tbinom{2 k}{k}^2+\frac{259}{20} \sum_{k=1}^a k^2 \tbinom{2 \
k}{k}^2
+\sum_{k=1}^a \tbinom{2 k}{k}^2+2 \sum_{k=1}^a k \tbinom{2 \
k}{k}^2,a]\\\vspace*{-0.06cm}
\COut \sum_{i_1=1}^a \tbinom{2 i_1}{i_1}{}^2-\sum_{i_1=1}^a \tbinom{2 \
i_1}{i_1}{}^2 i_1+\frac{1}{15} a (2 a+1)^2 (4 a+45) \tbinom{2 \
a}{a}^2\\
\end{Cmma}

\vspace*{-0.6cm}

\subsubsection*{Simplification with improved difference field theory}

\vspace*{-0.3cm}

The solution of Problem~EAR is obtained by calculating a set of algebraic independent sums (the generators of the \pisiSE-field) in which the occurring sums of the input expression can be rephrased. 
In order to guarantee that the output expression consists of sums and products that are simpler (or at least not more complicated) than the input expression, the generators of the \pisiSE-field must be constructed such that certain optimality criteria are fulfilled.
In short, we refine Problem~EAR using improved \pisiSE-difference field theory and enhanced algorithms for Problem~T. The most useful features of \texttt{SigmaReduce} can be summarized as follows.

\noindent$\bullet$ \textit{Atomic representation.} By default all sums are split into atomic parts (using partial fraction decomposition) and an algebraic independent representation of the arising sums and products is calculated. In addition,
\SigmaP\ outputs sums such that the denominators have minimal degrees w.r.t.\ the summation index (i.e., if possible, the denominator w.r.t.\ the summation index is linear). A typical example is

\begin{Cmma}
\CIn SigmaReduce[\sum_{k=1}^a \Big(\frac{-2+k}{10 (1+k^2)}+\frac{(1-4 k-2 k^2)S[1,k]}{10 (1+k^2) (2+2 k+k^2)}+\frac{(1-4 k-2 k^2)S[3,k]}{5 (1+k^2) (2+2 k+k^2)}\Big),a]\\\vspace*{-0.05cm}
\COut \frac{a^2+4 a+5}{10 (a^2+2 a+2)}S[1,a]-\frac{(a-1)(a+1)}{5(a^2+2a+2)}S[3,a]_3(a)-\tfrac{2}{5}\sum_{k=1}^a\frac{1}{k^2}\\
\end{Cmma}

\noindent This feature relies on algorithms refining those given in~\cite{Schneider:07d}; for the special case of rational sums see, e.g,~\cite{Abramov:71,Paule:95}. 
By default this refinement is activated; it can be switched off by using the option \texttt{SimpleSumRepresentation->False}. 

\medskip

\noindent$\bullet$ By default the following fundamental problem is solved:\\[0.1cm] 
\MyFrame{\textbf{Problem DOS: Depth Optimal Summation.} \textit{Given} a nested hypergeometric sum expression.
\textit{Find} an alternative representation of a nested hypergeometric sum expression whose nesting depth is minimal. Moreover, each sum cannot be expressed by a nested hypergeometric sum expression with lower depth.}\\[0.2cm]
The solution to this problem is possible by the enhanced difference field theory of depth-optimal \pisiSE-fields\index{difference field!\pisiSE-field!depth-optimal} and the underlying telescoping algorithms; see~\cite{Schneider:08c,Schneider:10b}. E.g., we can flatten the harmonic sum $S_{3,2,1}(a)$ of depth 3 to sums of depth$\leq 2$: 

\begin{Cmma}
\CIn SigmaReduce[\sum_{i=1}^a \frac{1}{i^3}\sum_{j=1}^{i}
\frac{1}{j^2}\sum_{k=1}^{j}\frac{1}{k},a]\\\vspace*{-0.1cm}
\COut \sum_{i_1=1}^a \frac1{i_1^5}\sum_{i_2=1}^{i_1} \frac{1}{i_2}+\Big(\sum_{i_1=1}^a \frac{1}{i_1^3}\Big)\Big(\sum_{i_1=1}^a \frac{1}{i_1^2}\sum_{i_2=1}^{i_1} \frac{1}{i_2}\Big)-\sum_{i_1=1}^a \frac{1}{i_1^2}\big(\sum_{i_2=1}^{i_1} \frac{1}{i_2^3}\big)\big(\sum_{i_2=1}^{i_1} \frac{1}{i_2}\Big)\\
\end{Cmma}

\noindent This depth-optimal \pisiSE-field theory yields various structural theorems~\cite{Schneider:10a}, i.e., gives a priori certain properties how the telescoping solution looks like. In particular, this leads to very efficient algorithms (for telescoping but also for creative telescoping and recurrence solving given below) where we could work with more than 500 sums in a depth-optimal \pisiSE-field.
The naive (and usually less efficient) \pisiSE-field approach is used with the option \texttt{SimplifyByExt$\to$None}.

\vspace*{-0.07cm}

\begin{example}
For the 2186 harmonic sums~\eqref{Equ:HarmonicSums} with weight $\sum_{i=1}^k|m_i|\leq7$ all algebraic relations are determined~\cite{ABS:13b}. More precisely, using their quasi-shuffle algebra the sums could be reduced by the \texttt{HarmonicSums} package~\cite{Ablinger:12} to 507 basis sums. Then using the algorithms above we showed that they are algebraic independent.
\end{example}

\vspace*{-0.07cm}


\smallskip

\noindent$\bullet$ \textit{Reducing the number of objects and the degrees in the summand.} The depth-optimal representation can be refined further as follows.\\[0.1cm] 
\textit{Given} a nested hypergeometric sum expression, \textit{find} an alternative sum representation such that for the outermost summands the number of occurring objects is as small as possible (more precisely, concerning a given tower of a \pisiSE-field the smallest subfield is searched in which the summand can be represented); see~\cite{Schneider:04a}. 

\smallskip

\noindent E.g., in the following example we can eliminate $S_1(k)$ from the summand:

\begin{Cmma}
\CIn SigmaReduce[\sum_{k=0}^a (-1)^k S_1(k){}^2 \binom{n}{k},a,SimplifyByExt\to DepthNumber]\\\vspace*{-0.1cm}
\COut -(a-n) \big(n^2 S_1(a){}^2+2 n S_1(a)+2\big) \frac{(-1)^a \binom{n}{a}}{n^3}-\frac{2}{n^2}-\frac{1}{n}\sum_{i_1=1}^a \frac{(-1)^{i_1}}{i_1}\binom{n}{i_1}\\
\end{Cmma}

\vspace*{-0.1cm}

\noindent Furthermore, one can calculate representations such that the degrees (w.r.t.\ the top extension of a \pisiSE-field) in the numerators and denominators of the summands are minimal~\cite{Schneider:07d}. For algorithms dealing with the product case we point to~\cite{Schneider:05c,Petkov:10}.

\subsection{Finding recurrence relations for definite sums}\label{Sec:RecurrenceFinding}

\textit{Given} a sum, say\footnote{\label{Footnote:IL}$L(m_1,\dots,m_u,n)$, $u\geq0$, stands for a linear combination of the $m_i$ and $n$ with integer coefficients.} $A(n)=\sum_{k=0}^{L(m_1,\dots,m_u,n)}F_n(k)$ where $F_n(k)$ is a nested hypergeometric sum expression depending on a discrete parameter $n$, \textit{find} polynomials $c_0(n),\dots,c_d(n)$ (not all zero) and an expression $h(n)$ in terms of sums that are simpler (see below) than the sum $A(n)$ such that the following linear recurrence holds:

\vspace*{-0.2cm}

\begin{equation}\label{Equ:RecForSum}
c_0(n)A(n)+\dots+c_d(n)A(n+d)=h(n).
\end{equation}

\smallskip

\noindent We treat this problem by the following variation of creative telescoping~\cite{Zeilberger:91}. 

\vspace*{0.1cm}

\noindent\MyFrame{\noindent\textbf{Problem CT: Creative telescoping (general paradigm).\index{creative telescoping!difference fields}}\\
\textit{Given} $d\in\NN$ and $F_n(k)$ such that $F_{n+i}(k)$ with $i\in\NN$ ($0\leq i\leq d$) can be written as nested hypergeometric sum expression.
\textit{Find} $\lambda\in\NN$, $c_0,\dots,c_d\in\KK(n)$, not all zero, and $G(k)$ such that for all $k\geq\lambda$ we have 
\begin{equation}\label{Equ:CreaSol}
c_0F_n(k)+c_1F_{n+1}(k)+\dots+c_dF_{n+d}(k)=\Shift_k G(k)-G(k)
\end{equation}
and such that the summands of the occurring sums in $G(k)$ are simpler (depending on the chosen strategy, see below) than $F_{n+i}(k)$; if this is not possible, return $\bot$.}

\vspace*{0.1cm}

The following solution relies on~\cite{Karr:81,Schneider:05a,Schneider:10c}.\\
\noindent\textit{Solution$^*$.} We consider the parameter $n$ as variable. Compute an ``appropriate'' polynomial \pisiSE-field $\dfield{\FF}{\sigma}$ over the constant field $\KK(n)$ as in Definition~\ref{Equ:PSPolyRing} with $R$ defined as in~\eqref{Def:PolyPiSi} together with an evaluation function $\fct{\ev}{R\times\NN}{\KK}$ in which one obtains explicitly $f_0,\dots,f_d\in R$ with $\lambda'\in\NN$ such that $\ev(f_i,k)=F_{n+i}(k)$ for all $k\geq\lambda'$; again we point to Footnote~\ref{Footnote:AlternatingSign}. 
Compute, if possible, a solution $c_0,\dots,c_d\in\KK(n)$ (not all zero) and $g\in R$ (or an extension of $\dfield{R}{\sigma}$ with an extended evaluation function $\ev$) such that

\vspace*{-0.4cm}

\begin{equation}\label{Equ:ParaTele}
c_0f_0+\dots+c_df_d=\sigma(g)-g
\end{equation}

\vspace*{-0.05cm}

\noindent holds; in addition we require that in $g$ the summands of the occurring sum extensions are simpler (depending on the chosen strategy, see below) than each of the given $f_i$.
If there is not such a solution, return $\bot$. Otherwise extract a nested hypergeometric sum expression $G(k)$ such that $G(k)=\ev(g,k)$ and compute $\lambda\in\NN$ such that~\eqref{Equ:CreaSol} holds for all $k\geq\lambda$; see Remark~\ref{Remark:PLDEGen} with $m=1$, $\alpha_1=1,\alpha_0=-1$. Then return $(c_0,\dots,c_d,G(k))$ and $\lambda$.\\[0.1cm]
\textit{Application.} Usually, one loops over $d=0,1,\dots$ until a solution for~\eqref{Equ:CreaSol} is found; for termination issues see Remark~1 on page~\pageref{Page:Remark:EMS}. Then summing~\eqref{Equ:CreaSol} over a valid range, e.g., from $\lambda$ to $a$, 
gives

\vspace*{-0.3cm}

\begin{equation}\label{Equ:RecFixedUpperBound}
c_0(n)\sum_{k=\lambda}^aF_n(k)+\dots+c_d(n)\sum_{k=\lambda}^aF_{n+d}(k)=G(a+1)-G(\lambda)
\end{equation}

\vspace*{-0.1cm}

\noindent where by construction the summands of the arising sums in $\bar{h}(a):=G(a+1)-G(\lambda)$ are simpler than $F_{n+i}(k)$. This implies that also the arising sums $\bar{h}(a)$ are simpler than $\sum_{k=\lambda}^aF_{n+i}(k)$. Note that so far $n$ is considered as an indeterminate. We remark that in many applications $n$ itself is an integer valued parameter and extra caution is necessary to avoid poles when summing up~\eqref{Equ:CreaSol}. Finally, when setting $a=L(m_1,\dots,m_u,n)$ in~\eqref{Equ:RecFixedUpperBound} (if $a=\infty$, a limit has to be performed) and taking care of missing summands yields~\eqref{Equ:RecForSum} for $A(n)$; see Example~\ref{Exp:Crea} for details.\\[0.1cm]
\textit{Proof certificate.} The correctness of~\eqref{Equ:RecForSum} for a given sum $A(n)$ is usually hard to prove. However, given the proof certificate $(c_0,\dots,c_d,G(k))$ it can be easily verified that~\eqref{Equ:CreaSol} holds within the required summation range. Then summing this equation over this range yields the verified result~\eqref{Equ:RecForSum}. 

\medskip

\noindent With \SigmaP\ one can calculate for $A_{-3}(n)=\texttt{SUM[n]}$ a recurrence\footnote{For a rigorous verification the proof certificate $(c_0,\dots,c_d,G(k))$ of~\eqref{Equ:RecFixedUpperBound} with $d=2$ is returned with the function call \texttt{CreativeTelescoping[mySum,n]}.} as follows:
\begin{Cmma}\CMLabel{ExpMMASum}
\CIn mySum=\sum_{k=0}^n(1-3(n-2k)S_1(k))\binom{n}{k}^{-3};\\
\end{Cmma}
\begin{Cmma}\CMLabel{ExpMMARec}
\CIn rec=GenerateRecurrence[mySum,n][[1]]\\
\COut (n+2)^4 (n+3)^2SUM[n]+(n+1)^3 (n+3)^2(2n+5)SUM[n+1]+(n+1)^3 (n+2)^3SUM[n+2]\newline
==\big(20 n^3+138 n^2+311 n+229\big) (n+1)^2
+6 (n+2)^2 (n+3)(2 n+5) (n+1)^3 S_1(n)\\
\end{Cmma}
\noindent The essential calculation steps are given in the following example.

\begin{example}$^*$\label{Exp:Crea}
Take $A_{-3}(n)=\sum_{k=0}^nF_n(k)$ with $F_n(k)=(1-3(n-2k)S_1(k))\binom{n}{k}^{-3}$. We calculate a recurrence for $A_{-3}(n)$ in $n$ by the techniques described above. First, we search for a solution of~\eqref{Equ:CreaSol} with $d=0$ (which amounts to telescoping). I.e., we construct the polynomial \pisiSE-field $\dfield{\QQ(n)(x)(b)(h)}{\sigma}$ and evaluation function $\fct{ev}{R\times\NN}{\QQ(n)}$ with $R:=\QQ(n)(x)[b,b^{-1}][h]$ as in Example~\ref{Exp:EvaluationDef}. There we take $f_0=(1-3(2n-x)h)b^{-3}\in R$ such that $\ev(f_0)=F_n(k)$ for all $k\geq0$. Unfortunately, our telescoping algorithm fails to find a $g\in R$ such that $\sigma(g)-g=f_0$ holds. So we try to find a solution of~\eqref{Equ:CreaSol} with $d=1$. Since $\binom{n+1}{k}=\frac{n+1}{n-k+1}\binom{n}{k}$, we can rephrase $F_{n+1}(k)$ by $f_1=(1-3(2n-x)h)\frac{(n-x+1)^3}{(n+1)^3}b^{-3}$, i.e., $\ev(f_1,k)=F_{n+1}(k)$ for all $k\geq0$. Then we activate the algorithm for Problem FPLDE and search for $c_0,c_1\in\QQ(n)$ (not both zero) and $g\in R$ such that~\eqref{Equ:ParaTele} holds with $d=1$. Again there is no solution. We continue our search and take $f_2=(1-3(2n-x)h)\frac{(n-x+1)^3(n-x+2)^3}{(n+1)^3(n+2)^3}b^{-3}$ with $\ev(f_2,k)=F_{n+2}(k)$ and look for $c_0,c_1,c_2\in\QQ(n)$ (not all zero) and $g\in R$ such that~\eqref{Equ:ParaTele} holds with $d=2$. This time our algorithm for Problem~FPLDE outputs $c_0=(n+2)^4 (n+3)^2$, $c_1=(n+1)^3 (n+3)^2(2n+5)$, $c_2=(n+1)^3 (n+2)^3$ and $g=(p_1(n,x)+p_2(n,x) h)b^{-3}$ for polynomials $p_1(n,x),p_2(n,x)\in\QQ[n,x]$. 
Hence we get~\eqref{Equ:CreaSol} with
$G(k)=\ev(g,k)=(p(n,k)+p_2(n,k)S_1(k))\binom{n}{k}^{-3}$. We emphasize that the correctness of~\eqref{Equ:CreaSol} for the given solution for all $k$ with $0\leq k\leq n$ can be verified easily. Finally, summing~\eqref{Equ:CreaSol} over $k$ from $0$ to $n$ one gets
\begin{multline*}
c_0(n)A_3(n)+c_1(n)(A_3(n+1)-F_{n+1}(n+1))\\
+c_2(n)(A_3(n+2)-F_{n+2}(n+1)-F_{n+2}(n+2))=G(n+1)-G(n);
\end{multline*}
moving the $F_{n+i}(n+j)$ terms to the right hand side gives the recurrence~\CmyOut{\ref{ExpMMARec}}. 
\end{example}

Note that creative telescoping is only a slight extension of telescoping, in particular, all the enhanced telescoping algorithms from Section~\ref{Sec:IndefiniteSummation} carry over to creative telescoping. In all variations, a polynomial \pisiSE-field $\dfield{\FF}{\sigma}$ (more precisely a depth-optimal \pisiSE-field~\cite{Schneider:05f} for efficiency reasons)
is constructed in which the summands $F_{n+i}(k)$ ($0\leq i\leq d$) can be expressed. Starting from there, the following tactics are most useful to search for a solution of~\eqref{Equ:ParaTele}. They are activated by using the option \texttt{SimplifyByExt->Mode} where \texttt{Mode} is chosen as follows.

\smallskip

\noindent$\bullet$ \textit{None:} The solution $G(k)$ is searched in $\dfield{\FF}{\sigma}$, i.e., only objects occurring in $F_{n+i}(k)$ are used. Here a special instance of FPLDE is solved; see Example~\ref{Exp:Crea}.

\smallskip

\noindent$\bullet$ \textit{MinDepth:} The solution $G(k)$ is searched in terms of sum extensions which are not more nested than the objects in $F_{n+i}(k)$ and which have minimal depth among all the possible choices~\cite{Schneider:08c}. This is the default option. 

\smallskip

\noindent$\bullet$ \textit{DepthNumber:} The solution is given in terms of sum extensions which are not more nested than $\sum_{k=0}^nF_{n+i}(k)$, however, if the nesting depth is the same, the number of the objects in the summands must be smaller than in $F_{n+i}(k)$. If such a recurrence exists, the machinery from~\cite{Schneider:01} computes it. Using this refined version for our example, one finds a recurrence of order 1 (instead of 2)

\vspace*{-0.5cm}

\begin{multline}\label{Equ:RecWithSum}
(n+1)^3A_{-3}(n+1)+(n+2)^3 A_{-3}(n)\\[-0.2cm]
=6 (n+2) (n+1)^3 S_1(n)+(7 n+13) (n+1)^2+3(n+2)^2 \sum_{i=0}^n (n-2 i)\binom{n}{i}^{-3}
\end{multline}

\vspace*{-0.4cm}

\noindent where the sum $E(n)=\sum_{i=0}^n(n-2 i)\binom{n}{i}^{-3}$ does not contain $S_1(i)$; it turns out that $E(n)=0$ (using again our tools) and the recurrence simplifies further to 
\begin{equation}\label{Equ:RecWithoutSum}
(n+1)^3A_{-3}(n+1)+(n+2)^3 A_{-3}(n)
=6 (n+2) (n+1)^3 S_1(n)+(7 n+13) (n+1)^2.
\end{equation}



\vspace*{-0.2cm}

\subsection{Solving recurrence relations}\label{Sec:RecurrenceSolving}

\vspace*{-0.2cm}

Next, we turn to recurrence solving in terms of nested hypergeometric sum expressions, i.e., expressions that evaluate to d'Alembertian sequence solutions\index{sequence!d'Alembertian}. 

\begin{example}
Given the recurrence \texttt{rec} in \CmyOut{\ref{ExpMMARec}} of  $A_{-3}(n)=\texttt{SUM[n]}$, all nested hypergeometric sum solutions are calculated with the following \SigmaP\ command:

\begin{Cmma}\CMLabel{ExpMMARecSol}
\CIn recSol=SolveRecurrence[rec,SUM[n]]\\
\COut \{\{0,-(-1)^n (n+1)^3\},\{0,(-1)^n \big(-S_1(n) (n+1)^3-(n+1)^2\big)\},\newline
\{1,6 (n+1) S_1(n)+(-1)^n \big(5 (n+1)^3 S_{-3}(n)-6 (n+1)^3 S_{-2,1}(n)\big)+1\}\}\\
\end{Cmma}

\noindent The output means that we calculated two linearly independent solutions $H_1(n)=-(-1)^n (n+1)^3$ and $H_2(n)=(-1)^n \big(-S_1(n) (n+1)^3-(n+1)^2\big)$  (for $n\geq0$) of the homogeneous version of the recurrence and a particular solution $P(n)=6 (n+1) S_1(n)+(-1)^n \big(5 (n+1)^3 S_{-3}(n)-6 (n+1)^3 S_{-2,1}(n)\big)+1$ (for $n\geq0$) of the recurrence itself; since the solutions are indefinite nested, the verification of the correctness can be verified easily by rational function arithmetic. Note that 
$\{c_1 H_1(n)+c_2H_2(n)+P(n)|c_1,c_2\in\QQ\}$
produces all sequence solutions whose entries are from $\QQ$. Since also $A_{-3}(n)$ is a solution of the recurrence, there is an element in $L$ that evaluates to $A_{-3}(n)$ for all $n\geq0$. 
Using, e.g., the first two initial values $A_{-3}(0)=1$ and $A_{-3}(1)=5$ the $c_1,c_2$ are uniquely determined: $c_1=c_2=0$. Thus we arrive at 
$A_{-3}(n)=P(n)$, i.e., we discovered and proved the identity~\eqref{Id:AM3} for $n\geq0$ (recall that we verified that both sides satisfy the same recurrence and that both sides agree with the first two initial values).
This last step is executed by taking \texttt{recSol} and $\texttt{mySum}=A_{-3}(n)$ (to get two initial values) as follows.

\begin{Cmma}
\CIn FindLinearCombination[recSol,mySum,n,2]\\
\COut 6 (n+1) S_1(n)+(-1)^n \big(5 (n+1)^3 S_{-3}(n)-6 (n+1)^3 S_{-2,1}(n)\big)+1\\
\end{Cmma}
\end{example}

\noindent In general, \SigmaP\ can solve the following problem~\cite{Norlund1924,Abramov:94,Schneider:01}.

\medskip

\noindent\MyFrame{\noindent\textbf{Problem RS: Recurrence solving.}\\
\textit{Given} polynomials $a_0(n),\dots,a_m(n)\in\KK(n)$ and a nested hypergeometric sum expression $f(n)$. \textit{Find} the full solution set of the $m$th-order linear recurrence
\begin{equation}\label{Equ:RecSpec}
a_0(n)G(n)+\dots+a_m(n)G(n+m)=f(n)
\end{equation}
in terms of nested hypergeometric sum expressions. I.e., return $\bot$ if there is no particular solution. Otherwise, find $\lambda\in\NN$ and nested hypergeometric sum expressions $((1,P(n)),(0,H_1(n)),\dots,(0,H_l(n)))$ where $P(n)$ is a particular solution and $H_1(n),\dots H_l(n)$ are solutions of the homogeneous version of~\eqref{Equ:RecSpec} for $n\geq\lambda$; the sequences (in $S(\KK)$) produced by $H_1(n),\dots H_l(n)$ are linearly independent. In addition, all
sequences $(G(n))_{n\geq0}\in\KK^{\NN}$, that are solutions of~\eqref{Equ:RecSpec} for all $n\geq\lambda$ and that can be given by nested hypergeometric sum expressions, can be also produced by 

\vspace*{-0.3cm}

\begin{equation}\label{Equ:SolSpace}
L=\{P(n)+c_1H_1(n)+\dots c_l H_l(n)|c_i\in\KK\}
\end{equation}
starting from $n\geq\lambda$.
}

\vspace*{0.1cm}
The following solution relies on~\cite{Karr:81,Petkov:92,Schneider:01,Schneider:05a,Schneider:10c}.\\
\noindent\textit{Solution$^*$.} 
Construct a polynomial \pisiSE-field $\dfield{\FF}{\sigma}$ as in Definition~\ref{Equ:PSPolyRing} with $R$ defined as in~\eqref{Def:PolyPiSi} together with an evaluation function $\fct{\ev}{R\times\NN}{\KK}$ in which one obtains explicitly a $\Phi\in R$ with $\lambda'\in\NN$ such that $\ev(\Phi,n)=f(n)$ for all $n\geq\lambda'$; again Footnote~\ref{Footnote:AlternatingSign} applies.
In other words, with $\alpha_i:=a_i(x)\in\KK(x)$ we can reformulate~\eqref{Equ:RecSpec} with
\begin{equation}\label{Equ:LinDiffEqu}
\alpha_0g+\alpha_1\sigma(g)+\dots+\alpha_m\sigma^m(g)=\Phi.
\end{equation}
Factorize the homogeneous recurrence (written as linear operator) as much as possible in linear right factors using Hyper~\cite{Petkov:92}. Each linear factor describes a hypergeometric solution which is adjoined to our \pisiSE-field (see Problem~RP); for simplicity we exclude the possible case that $(-1)^n$ is needed for this task. Applying Algorithm~\cite[Alg.~4.5.3]{Schneider:01} to this recurrence returns the output $((1,p),(0,h_1),\dots,(0,h_l))$ in a polynomial \pisiSE-field $\dfield{\set E}{\sigma}$ that contains $\dfield{\FF}{\sigma}$ with the following property: $p$ is a particular solution of~\eqref{Equ:LinDiffEqu} and the $h_i$ are $l$ linearly independent solutions of the homogeneous version of~\eqref{Equ:LinDiffEqu}. We omit details here and remark only that it is crucial to solve~\eqref{Equ:PLDE} as subproblem.  Then extend the evaluation function from $\FF$ to $\set E$ (we are free to choose appropriate lower bounds and constants of the sums/products), and let $H_i(n)$ ($1\leq i\leq l$) and $P(n)$ be the nested hypergeometric sum expressions that define the evaluations $\ev(h_i,n)$ and $\ev(p,n)$, respectively. Compute $\lambda$ such that 
the $H_i(n)$ are solutions of the homogeneous version and $P(n)$ is a particular solution of~\eqref{Equ:RecSpec} for all $n\geq\lambda$; see Remark~\ref{Remark:PLDEGen} with $d=0$, $f_0:=\Phi$.\\[0.1cm]
\normalsize
\textit{Remark.} (1) If one computes $m=l$ linearly independent solutions plus a particular solution, the set~\eqref{Equ:SolSpace} gives all solutions. If this is not the case, the completeness of the method, i.e., that no solution in terms of nested hypergeometric sum expressions is missed, needs further justification: it can be deduced from \cite[Cor.~4.5.2]{Schneider:01} and Remark~\ref{Remark:PLDEGen}; for deep insight and alternative proofs see~\cite{Singer:99} and~\cite{Petkov:2013}.\\
(2) The derived solutions are highly nested: For each additional solution one needs one extra indefinite sum on top. In most examples the simplification of these solutions (see Subsection~\ref{Sec:IndefiniteSummation}) is the most challenging task; see, e.g., Example~\ref{Exp:QCDSolving}.\\
(3) Since the solutions are indefinite nested, the shifted versions can be expressed by the non-shifted versions. Using this property and considering the sums and products as variables, the correctness can be verified by rational function arithmetic.\\
(4) Also the $a_i(n)$ in~\eqref{Equ:LinDiffEqu} can be from a \pisiSE-field and one can factorize the difference operator in linear factors; this is based on work by~\cite{Bron:00,Schneider:05a,ABPS:13}.

\begin{example}$^*$
We construct the polynomial \pisiSE-field $\dfield{\QQ(x)(h)}{\sigma}$ with $\sigma(x)=x+1$ and $\sigma(h)=h+\frac{1}{x+1}$ over $\QQ$ and interpret the elements with the evaluation function $\fct{\ev}{\QQ(x)[h]\times\NN}{\QQ(x)[h]}$ canonically defined by~\eqref{Equ:EvRat} and $\ev(h,n)=S_1(n)$. In this way, we can reformulate the recurrence~\CmyOut{\ref{ExpMMARec}} with

\vspace*{-0.7cm}

\begin{multline}\label{Equ:DFEquExp}
(x+2)^4 (x+3)^2\,g+(x+1)^3 (x+3)^2(2x+5)\sigma(g)+(x+1)^3 (x+2)^3\sigma^2(g)=\\
\big(20 x^3+138 x^2+311 x+229\big) (x+1)^2
+6 (x+2)^2 (x+3)(2 x+5) (x+1)^3 h.
\end{multline}

\vspace*{-0.1cm}

\noindent Then we execute the recurrence solver in this \pisiSE-field and get as output the difference ring $\dfield{\QQ(x)[m][h][s][H]}{\sigma}$ with $\sigma(m)=-m$ where $m^2=1$,
$\sigma(s)=s+\frac{-m}{(x+1)^3}$ and
$\sigma(H)=H+\tfrac{-m(h+\frac{1}{x+1})}{(x+1)^2}$ such that $\Dconst{\QQ(x)[m][h][H]}{\sigma}=\QQ$. There it returns the linearly independent solutions
$h_1=m (x+1)^3$
and $h_2=m\big(h (x+1)^3+(x+1)^2)$ of the homogeneous version of~\eqref{Equ:DFEquExp} and the particular solution
$p= 6 (x+1) h+m\big(5 (x+1)^3 s-6 (x+1)^3 H\big)+1$ of~\eqref{Equ:DFEquExp} itself. Note that the solutions (coming from the factorization of the recurrence) have been simplified already using the technologies presented in Section~\ref{Sec:IndefiniteSummation}. Finally, we extend the evaluation function from $\QQ(x)[h]$ to $\QQ(x)[m][h][H]$ by
$\ev(m,n)=(-1)^n$,
$\ev(s,n)=S_{-3}(n)$ and
$\ev(H,n)=S_{-2,1}(n)$. This choice yields
$H_1(n)=\ev(h_1,n)$,
$H_2(n)=\ev(h_2,n)$
and $P(n)=\ev(p,n)$ as given in~\CmyOut{\ref{ExpMMARecSol}}.
\end{example}

\begin{example}\label{Exp:QCDSolving}
In~\cite{BKKS:09} (see also~\cite{Kauers:13}) recurrences are guessed with minimal order that contain as solutions the massless
Wilson coefficients to 3-loop order for individual color coefficients~\cite{Moch:04}. Afterwards the recurrences have been solved. The largest recurrence of order 35 could be factorized completely into linear factors in about 1 day. This yields 35 linearly independent solutions in terms of sums up to nesting depth 34. Then their simplifications in terms of harmonic sums took 5 days.
\end{example}

\section{Simplification of multiple sums with \texttt{EvaluateMultiSums}}\label{Sec:EMS}

In Section~\ref{Sec:SumTechniques} we transformed the definite sum $A_{-3}(n)$ to a nested hypergeometric sum expression given in~\eqref{Id:AM3} by calculating a recurrence and solving it. Applying this tactic iteratively leads to a successful method to transform certain classes of definite multiple sums\index{summation!multiple sum} to nested hypergeometric sum expressions. Consider, e.g.,

\vspace*{-0.5cm}

\begin{equation}\label{Equ:EMSInput}
F(n):=\sum_{j=0}^{n-2} (-j+n-2)! \overbrace{\sum_{r=0}^{j+1} \frac{(-1)^r \binom{j+1}{r} r!}{(-j+n+r)!}\underbrace{\sum_{s=0}^{-j+n+r-2} \frac{(-1)^s \binom{-j+n+r-2}{s}}{(n-s)
(s+1)}}_{=:F_0(n,j,r)}}^{=:F_1(n,j)}
\end{equation}
which arose in QCD calculations needed in~\cite{HYP2}; see also~\cite{Schneider:10d}. We zoom into the sum $F_0(n,j,r)$, a definite sum over a hypergeometric sequence. Calculating a recurrence, solving the recurrence, and combining the solutions leads to the simplification

\vspace*{-0.4cm}

\begin{equation*}
F_0(n,j,r)=\frac{1}{(n+1) (-j+n+r-1)}
+\frac{(-1)^n (j+1)!(-j+n-1)_r}{(n-1) n (n+1) (-j-1)_r (2-n)_j}.
\end{equation*}

\vspace*{-0.1cm}

\noindent This closed form could be also derived by hypergeometric summation~\cite{AequalB}: the double sum $F_1(n,j)$ turns out to be a single sum. Next, finding a recurrence for this sum and solving the recurrence lead to a nested hypergeometric sum expression w.r.t.\ $j$:
\small
\begin{equation*}
F_1(n,j)=(-1)^j(j+1)!
\Big[\frac{1}{n!}\Big(\tfrac{(-1)^n
(j+2)}{(n+1)^2 (-j+n-1)}+\tfrac{n^2+1}{(n-1) n (n+1)^2}\Big)\\
+\frac{1}{n+1} \sum_{i=1}^j \tfrac{(-1)^{i}}{\big(n-i\big)!
\big(i+1\big)! \big(n-i-1\big)}\Big].
\end{equation*}
\normalsize
In other words, $F(n)$ can be written as a definite sum where the summand is a nested hypergeometric sum expression. Therefore we are again in the position to apply our technologies from Section~\ref{Sec:SumTechniques}. Computing a recurrence and solving it yields
\begin{equation*}
F(n)=\frac{-n^2-n-1}{n^2 (n+1)^3}+\frac{(-1)^n \big(n^2+n+1\big)}{n^2 (n+1)^3}
+\frac{S_1(n)}{(n+1)^2}-\frac{S_2(n)}{n+1}-\frac{2 S_{-2}(n)}{n+1}.
\end{equation*}
Summarizing, we transformed a definite nested sum from inside to outside to a nested hypergeometric sum expression. 
More generally, we deal with the following

\medskip

\noindent\MyFrame{\noindent\textbf{Problem EMS: EvaluateMultiSum.}
\textit{Given} a definite multiple sum

\vspace*{-0.2cm}

\begin{equation}\label{Equ:EMSInputSum}
F(\vect{m},n)=\sum_{k=l}^{L_0(\vect{m},n)}\overbrace{\sum_{k_1=l_1}^{L_1(\vect{m},n,k)} ... \sum_{k_v=l_v}^{L_v(\vect{m},n,k,k_1, ..., k_{v-1})}
\overline{f}(\vect{m},n,k,k_1,\dots,k_v)}^{f(\vect{m},n,k)}
\end{equation}
\noindent with a nested hypergeometric sum expression $\overline{f}$ w.r.t.\ $k_v$, integer parameters $n$ and $\vect{m}=(m_1\dots,m_r)$, and $L_i(\dots)$ being integer linear (see Footnote~\ref{Footnote:IL}) or $\infty$.\\[0.1cm]
\textit{Find} $\lambda\in\NN$ and a nested hypergeometric sum expression $\bar{F}(\vect{m},n)$ such that $F(\vect{m},n)=\bar{F}(\vect{m},n)$ for $n\geq\lambda$.}

\medskip

\noindent\textit{Method.} Apply the techniques of Section~\ref{Sec:SumTechniques} recursively as follows~\cite{Schneider:12a}.
\begin{enumerate}
\item Transform the outermost summand $f(\vect{m},n,k)$ to a nested hypergeometric sum expression w.r.t.\ $k$ by applying the proposed method recursively to all the arising definite sums (i.e.,  the parameter vector $\vect{m}$ is replaced by $(\vect{m},n)$ and the role of $n$ is $k$). Note that the sums in $f$ are simpler than $F(\vect{m},n)$ (one definite sum less). If the summand $f$ is free of sums, nothing has to be done.

\item Solve Problem CT: Compute a recurrence~\eqref{Equ:RecForSum} for the sum $A(n)=F(\vect{m},n)$; if this fails, ABORT.
If successful (say it is of order $o$), the right hand side might be again an expression in terms of definite sums, but their summands are simpler than $f$ (see, e.g., the recurrence~\eqref{Equ:RecWithSum}). Apply the method recursively to these sums such that the right hand side is transformed to a nested hypergeometric sum expression w.r.t.\ $n$ (see, e.g., recurrence~\eqref{Equ:RecWithoutSum}).

\item Solve Problem RS: Compute all nested hypergeometric sum solutions of the recurrence~\eqref{Equ:RecForSum} and simplify the solutions using the techniques from~Section~\ref{Sec:IndefiniteSummation}.

\item Compute $o$ initial values, i.e., specialize the parameter $n$ to appropriate values from $\set N$, say $n=l,l+1,\dots,l+o-1$, and apply the method recursively to the arising sums where $m_1$ takes over the role of $n$ and the remaining parameters are $(m_2,\dots,m_r)$. If no parameter is left, the expression is a constant. It is usually from $\QQ$ (if no sum is left) or it simplifies, e.g., to multiple zeta values~\cite{MZV} or infinite versions of $S$--sums~\cite{ABS:13} and cyclotomic sums~\cite{ABS:11}.

\item Try to combine the solutions to find a nested hypergeometric sum expression w.r.t.\ $n$ of $F(\vect{m},n)$. If this fails, ABORT. Otherwise return the solution.
\end{enumerate}

\noindent\textit{Remark.}\label{Page:Remark:EMS} (1) The \textit{existence} of a recurrence in Step~2 is guaranteed in many cases (in particular for sums coming from Feynman integrals~\cite{BKSF:12}) by using arguments, e.g., form~\cite{Wilf:92,AequalB,Wegschaider,AZ:06}. Here often
computation issues are a bottleneck. Usually, we succeed in finding recurrences when $f$ consists of up to 100 nested hypergeometric sums. If $f$ is more complicated (or if it seems appropriate), the sum is split into several parts and the method is applied to each sum separately.\\
(2) \textit{Termination:} The method is applied recursively to sums which are always simpler than the original sum (less summation quantifiers, less parameters,  or less objects in the summand). Hence eventually one arrives at the base case.\\
(3) \textit{Success:} If the method does not abort in one of the executions of step~2 or step~5, it terminates and outputs a nested hypergeometric sum expression w.r.t.\ $n$. Note that finding not sufficiently many solutions of a given recurrence in step~5 is the main reason why the method might fail. For general multiple sums this failure would happen all over. However, e.g, in the context of Feynman integrals, the recurrence is usually completely solvable (i.e., we find $m$ linearly independent solutions of the homogeneous version of~\eqref{Equ:RecForSum} and one particular solution of the recurrence itself).

\smallskip

We emphasize that 3--loop Feynman integrals with at most 1 mass~\cite{BKSF:12} can be transformed to multiple sums and that the simplification of these sums is covered exactly by Problem EMS. The described method is implemented in the following new package which uses the summation algorithms in \SigmaP:

\vspace*{-0.1cm}

\begin{Cmma}
\CIn << EvaluateMultiSums.m \\
\CPrint EvaluateMultiSums\; by\; Carsten\;\; Schneider -- \copyright\ RISC\\
\end{Cmma}

\vspace*{-0.3cm}

\noindent In addition it uses (some of the many) functions from J. Ablinger's package \texttt{HarmonicSums}~\cite{Bluemlein:99,Vermaseren:99,Bluemlein2009a,ABS:11,Ablinger:12,ABS:13} to transform --if possible-- the arising indefinite sums to harmonic sums, $S$-sums, cyclotomic sums or their infinite versions, to find algebraic relations among these sums, and to calculate asymptotic expansions of these sums for limit computations (this is needed if upper bounds in~\eqref{Equ:EMSInputSum} are $\infty$).

\vspace*{-0.1cm}

\begin{Cmma}
\CIn << HarmonicSums.m \\
\CPrint HarmonicSums\; by\; Jakob\;\; Ablinger -- \copyright\ RISC\\
\end{Cmma}

\vspace*{-0.3cm}

\noindent Then inserting the summand with the summation ranges of~\eqref{Equ:EMSInput} and the information that there is the extra integer parameter $n$ with $2\leq n\leq\infty$ we can activate the simplification of the sum~\eqref{Equ:EMSInput} to a nested hypergeometric sum expression as follows.

\begin{Cmma}
\CIn EvaluateMultiSum[\tfrac{(-j+n-2)!(-1)^{r+s} \binom{j+1}{r} r!}{(-j+n+r)!}\tfrac{\binom{-j+n+r-2}{s}}{(n-s)
(s+1)},\newline
\hspace*{3cm}\{\{s,0,-j+n+r-2\},\{r,0,j+1\},\{j,0,n-2\}\},\{n\},\{2\},\{\infty\}]\\
\COut \frac{-n^2-n-1}{n^2 (n+1)^3}+\frac{(-1)^n \big(n^2+n+1\big)}{n^2 (n+1)^3}
+\frac{S_1(n)}{(n+1)^2}-\frac{S_2(n)}{n+1}-\frac{2 S_{-2}(n)}{n+1}\\
\end{Cmma}

\noindent Similarly, we can calculate the simplification given in identity~\eqref{Id:AM4}:

\begin{Cmma}
\CIn EvaluateMultiSum[
\big(1-4(n-2k)S_1(k)\big)\binom{n}{k}^{-4},\{\{k,0,n\}\},\{n\},\{0\},\{\infty\}]\\
\COut\frac{(10 (n+1) S_1(n)+3)
   (n+1)}{2 n+3}
+\frac{(-1)^n\binom{2n}{n}^{-1}(n+1)^5}{(4 n
   (n+2)+3)}\left(\frac{7}{2} \sum
   _{i=1}^n \frac{(-1)^i \binom{2i}{i}}{i^3}-5 \sum _{i=1}^n
   \frac{(-1)^i \binom{2i}{i} S_1(i)}{i^2}\right).\\
\end{Cmma}


\noindent As mentioned already above, the multiple sums coming from many 2--loop and 3-loop Feynman integrals fit into the input class of the package \texttt{EvaluateMulti}\-\texttt{Sums}.  Here two extremes occurred: In~\cite{HYP2,BHKS:13}
about a million multiple sums (mostly triple and quadruple sums) were simplified.
Using the package \texttt{Sum\-Production}~\cite{Schneider:12a} we merged the sums to several 100 basis sums where each of the summands required up to 20 MB memory. The other extreme are sums whose summands are in compact size, but the number of summations is large; one of the most complicated input sums from~\cite{BHKS:13} is, e.g.,~\eqref{Equ:QCDSum}. In both setups the transformed summands during the \texttt{EvaluateMultiSum} method became rather large containing complicated nested hypergeometric sums. Only in the last step these nasty sums vanished and the expected nice result popped up; note that already for the transformation of the sum~\eqref{Equ:EMSInput} this effect is visible. Summarizing, the summation algorithms based on enhanced difference field theory, presented in this article, were indispensable to master the challenging calculations as given, e.g., in~\cite{Schneider:08e,HYP2,ABHKSW:12,BHKS:13}.

\begin{acknowledgement}
Supported by the Austrian Science Fund (FWF) grants P20347-N18 and SFB F50 (F5009-N15) and
by the EU Network {\sf LHCPhenoNet} PITN-GA-2010-264564.
\end{acknowledgement}
%

%
%


\end{document}